\documentclass{article}
\usepackage[utf8]{inputenc}
\usepackage[margin=3cm]{geometry}
\usepackage{natbib}
\usepackage{amsmath, amssymb, graphicx, psfrag}
\usepackage{algorithm, algorithmic}
\usepackage{url}

\title{Ensemble MCMC: Accelerating Pseudo-Marginal MCMC for State Space Models using the Ensemble Kalman Filter}
\author{Christopher Drovandi\footnote{Ordering of authors is alphabetical} \\ School of Mathematical Sciences \\ Australian Centre of Excellence for Mathematical and Statistical Fronters \\ Queensland University of Technology, Australia \\ c.drovandi@qut.edu.au \\ \\
	Richard G Everitt \\ Department of Mathematics and Statistics \\ Univeristy of Reading, UK \\ r.g.everitt@reading.ac.uk \\ \\
	Andrew Golightly \\ School of Mathematics, Statistics and Physics \\ Newcastle University, UK \\ andrew.golightly@ncl.ac.uk \\ \\
	Dennis Prangle \\ School of Mathematics, Statistics and Physics \\ Newcastle University, UK \\ dennis.prangle@ncl.ac.uk  }
\date{\today}

\begin{document}
	
	\setlength{\parindent}{0pc}
	\setlength{\parskip}{1ex}
	
	\maketitle
	\begin{abstract}
		Particle Markov chain Monte Carlo (pMCMC) is now a popular method for performing Bayesian statistical inference on challenging state space models (SSMs) with unknown static parameters.  It uses a particle filter (PF) at each iteration of an MCMC algorithm to unbiasedly estimate the likelihood for a given static parameter value.  However, pMCMC can be computationally intensive when a large number of particles in the PF is required, such as when the data is highly informative, the model is misspecified and/or the time series is long.  In this paper we exploit the ensemble Kalman filter (EnKF) developed in the data assimilation literature to speed up pMCMC.  We replace the unbiased PF likelihood with the biased EnKF likelihood estimate within MCMC to sample over the space of the static parameter.  On a wide class of different non-linear SSM models, we demonstrate that our new ensemble MCMC (eMCMC) method can significantly reduce the computational cost whilst maintaining reasonable accuracy.  We also propose several extensions of the vanilla eMCMC algorithm to further improve computational efficiency.  Computer code to implement our methods on all the examples can be downloaded from \url{https://github.com/cdrovandi/Ensemble-MCMC}.	 
	\end{abstract}

	%% *** Frontmatter *** 
	
\section{Introduction}\label{sec:intro}

%ADD MOTIVATION FOR INTEREST IN STATE SPACE MODELS?

Particle Markov chain Monte Carlo (pMCMC, \citealp{Andrieu2010}) is now a popular method for performing Bayesian statistical inference on challenging state space models (SSMs) with unknown static parameters. The appeal of particle MCMC is that it is a pseudo-marginal method \citep{andrieu+r09}, which attempts to mimic the ideal sampler that proposes directly over the space of the static parameters and integrates out the hidden states. Furthermore it is an \emph{exact approximation}, exactly targeting the true posterior distribution.

Each static parameter proposal in pMCMC is evaluated using a particle filter \citep{Gordon1993}. Particle filters were originally proposed to solve the state space \emph{filtering} problem: inferring the state parameters at a given time under known static parameters. To do so they propagate a set of particles through the state space model, and use a weighting and resampling process to concentrate on the particles with significant posterior weights. Using particle filters in pMCMC is costly. Firstly, each particle filter involves processing the entire dataset. Secondly, a particle filter can require a large number of particles, especially when the data is highly informative and/or the model is misspecified. This is because there must be enough particles to randomly propagate forwards to produce good matches to unlikely data. Thus, despite the popularity of particle MCMC, it is generally a highly computationally intensive method.

Data assimilation (DA) is a field of research originating in the geosciences, initially based on the problem of numerical weather prediction. The task most commonly addressed in this field is the estimation of the state of a dynamical system, based on a dynamic model (usually a system of partial differential equations) and noisy and/or indirect measurements of this state. In this paper we take inspiration from the DA literature to propose a new approach to estimating the posterior distribution of static parameters in SSMs.

The field of DA has evolved in parallel to other fields in which SSMs play an important role, such as target tracking, economics and statistical ecology. The distinguishing feature of problems in DA is the large dimension of the state space. For example, in numerical weather prediction the state space consists of a representation of the state of the atmosphere across the globe, which for modern applications can have dimension $d_{x}$ of the order of $10^{9}$ \citep{vanLeeuwen2015}. The traditional approach to estimating the dynamic state in DA is to use approaches that solely estimate the mode of the state posterior (e.g.\ 4DVar) or Kalman filters that make use of approximations so as to avoid storing the full state covariance, whose size scales quadratically in the state dimension. Such methods have huge practical importance and are still deployed in DA applications, but more recent research has focussed on methods that improve the accuracy of state estimation when using nonlinear dynamics. As in other fields where this the case, particle filters are an important methodology.

Particle filters are not the usual method of choice in DA. The reason is their degeneracy when used on states of high dimension \citep{Snyder2008}. This degeneracy arises due to the limitations of importance sampling in high dimensions: the variance of importance sampling estimators depends on the distance between the target and proposal distributions, and this distance grows with dimension such that the variance is only controlled by using a number of importance points that is exponential in the dimension (see \citealp{Agapiou2017} for a  review). To combat this degeneracy, the approach usually taken in the particle filtering literature is to introduce diversity into the sample though using MCMC updates of the state \citep{Beskos2014}. However, in many problems in DA, MCMC updates are often not available due to the use of an intractable dynamic model, and where available may have a low acceptance rate. An alternative means of maintaining diversity is given by the ensemble Kalman filter (EnKF) (\citealp{evensen1994}; see \citealp{Katzfuss2016} for a tutorial). This approach propagates a set of particles (often refered to as ``ensemble members'') through the dynamic model in the same way as the bootstrap particle filter, but instead uses these particles to approximate a Gaussian representation of the state distribution. This method approximates the Kalman filter and provides a means to avoid storing and manipulating the state covariance matrix. Further, it has also been shown to perform well when applied to non-linear dynamic models. Its performance is often superior to the particle filter in cases where particle filter suffers from degeneracy, including but not limited to the case of a state space of high dimensions.

Methods from DA have also been applied to the situation of inferring an unknown static parameter simultaneously with the state. The standard approach is to augment the state vector with the static parameter, then to apply one of the previously mentioned filters to this augmented state (see, for example, \citet{Evensen2007}. In this case, the EnKF assumes that both parameters and states follow a linear Gaussian state space model. When this assumption is unreasonable, another approach is to combine the EnKF likelihood with a particle representation of the static parameter \citep{Stroud2018,Katzfuss2019}. To mitigate degeneracy, the static parameter is allowed to dynamically vary, by adding Gaussian noise to each parameter particle. This step can be further refined via the kernel resampling strategy of 
\citet{Liu2001}. However, additional tuning parameters must be specified (e.g. to control the smoothness of the kernel) and the particle approximation can be sensitive to these choices, and in in particular, the number of particles used \citep{vieira2016}.

%A different, approximate, approach to reduce computation associated with complex state space models is the ensemble Kalman filter (EnKF, . The ensemble Kalman filter is fast as it processes the data only a single time and can require fewer particles; referred to as ensemble members in the EnKF context. Further, the updates are made explicitly conditional on the next data point, and is thus better than particle filtering at handling informative or surprising data However, it approximately propagates the uncertainty in both the static and state parameter via a Gaussian assumption. This assumption can result in a crude approximation of the static parameter posterior.

In this paper we propose a new method, called ensemble MCMC (eMCMC), which is a compromise between pMCMC and EnKF. It can be viewed as a pMCMC algorithm, where each use of a particle filter is replaced with using the EnKF (with fixed static parameters). The motiviation is to reduce the number of particles/members required and to cope better with informative or surprising data. Thus we reduce the computational cost relative to pMCMC while improving the accuracy relative to EnKF with time-varying static parameters. 

Moreover, we propose several extensions of eMCMC to further improve computational efficiency and reduce the bias in the EnKF estimate of the likelihood.  Some of the extensions may be of interest to the data assimilation community more generally.  The basic idea underlying eMCMC is also suggested in \citet{Katzfuss2019}.  However, \citet{Katzfuss2019} consider only a simple example.  Here we demonstrate that the method can be successful on a wider variety of more challenging applications and also develop several extensions just discussed. 

The rest of this article is structured as follows.  In Section \ref{sec:background} we provide the necessary background on state space models, pMCMC and EnKF to understand our method.  Our ensemble MCMC approach together with extensions is described in Section \ref{sec:method}.  Section \ref{sec:results} shows the results of our approach on a wide class of different models.  We discuss limitations, further extensions and possible future research in Section \ref{sec:discussion}.

\section{Background}\label{sec:background}

This section describes relevant existing work.
Section \ref{subsec:ssm} defines state space models.
Section \ref{subsec:pmcmc} describes pseudo-marginal Metropolis Hastings and the bootstrap particle filter, which can be used to perform inference for these models.
Section \ref{subsec:enkf} introduces the EnKF.

\subsection{State Space Models} \label{subsec:ssm}

A state space model is a model for sequential data.
It introduces a \emph{Markov chain} of latent \emph{states} $x_1, \ldots, x_T$.
Independent noisy observations $y_t$ are available that depend on the state 
$x_t$.
Let $x$ denote the collection of all latent states and $y$ the collection of all observations.
The model can be defined using an \emph{evolution} distribution for $x_{t+1} | x_t, \theta$
and an \emph{observation} distribution $y_t | x_t, \theta$.
Here $\theta$ is a vector of parameters controlling the model's behaviour.
We also specify a distribution for an initial state $x_0$.
%Could add dependence on parameters if needed for any examples
For more background on state space models see for example \cite{Sarkka2013}.

Throughout the paper we will make some standard assumptions about state space models.
We will assume that each $x_t$ and $y_t$ are random vectors with support $\mathbb{R}^{d_x}$ and $\mathbb{R}^{d_y}$ respectively.
In this section we assume the distributions above -- evolution, observation and initial state -- have densities $p(x_{t+1} | x_t, \theta), p(y_t | x_t,\theta)$ and $p(x_0)$.
The material immediately generalises to the case where some or all of these distributions have probability mass functions instead (by interpreting these as densities with respect to counting measure). 
This is required in several of our examples.  A point mass can be used for the initial state distribution if the initial state is known.

As we shall see, the EnKF is restricted to certain observation models.
Hence in this paper we focus on one particular case,
\begin{equation} \label{eq:Gaussian observation}
y_t | x_t \sim \mathcal{N}(P x_t, S)
\end{equation}
where $P$ is a $d_y \times d_x$ matrix and $S$ is a variance matrix, possibly a function of $\theta$.
(We assume conditional independence of the $y_t$'s given $x$ and $\theta$.)
The EnKF can also be used where $P$ is replaced by a time dependent matrix $P_t$.
The case where $P = I$ gives a \emph{complete observation} regime, in the sense that all components of $x_t$ have a corresponding noisy observation.
In contrast, a \emph{partial observation} regime only allows observation of a subset of the components
e.g.~by taking $P$ to be a projection matrix.

In practice we may wish to model states at a finer time discretisation than that at which the observation data is available.
For example, consider the case where we only have observations $y_t$ at $t=k,2k,\ldots,kL$.
This can easily be converted into the framework described above, by defining a state space model with $x^*_\tau = x_{\tau k}$ and $y^*_\tau = y_{\tau k}$ for $\tau=1,2,\ldots,L$.

The joint density of the latent states and observations in a state space model is:
\begin{equation} \label{eq:SSM joint}
p(x, y | \theta) = p(x_0) \prod_{t=1}^T \big[ p(x_t | x_{t-1}, \theta) p(y_t | x_t, \theta) \big].
\end{equation}
The likelihood can be found by marginalisation i.e.~integrating out the latent states $x$,
\begin{equation} \label{eq:SSM likelihood}
L(\theta) = \int p(x_0) \prod_{t=1}^T \big[ p(x_t | x_{t-1}, \theta) p(y_t | x_t, \theta) \big] dx.
\end{equation}
(If there is an observation $y_0$, a factor $p(y_0 | x_0, \theta)$ can easily be included in \eqref{eq:SSM joint} and \eqref{eq:SSM likelihood}.)

Bayesian inference assigns a prior $p(\theta)$ to the parameters and targets the posterior
$p(\theta | y) \propto p(\theta) L(\theta)$.
The likelihood $L(\theta)$ typically cannot be evaluated as it is a high dimensional integral.
One strategy to perform inference is to instead consider an augmented target density (often of interest in its own right), the joint posterior $p(\theta, x | y) \propto p(\theta) p(x, y | \theta)$.
The posterior for $\theta$ can then be obtained by marginalisation.

\subsection{Pseudo-marginal MCMC, particle filters, and particle MCMC} \label{subsec:pmcmc}

Monte Carlo algorithms are designed to sample from a target distribution, often a Bayesian posterior distribution.
Markov chain Monte Carlo (MCMC) does so using a Markov chain which converges to the target distribution in the long run.
Performing each update in MCMC typically requires likelihood calculations, which are not possible for models with intractable likelihoods.
However it is often possible to produce unbiased likelihood estimates.
Algorithm \ref{alg:PMMH}, \emph{pseudo-marginal Metropolis Hastings} (PMMH) \citep{andrieu+r09}, makes use of these to perform parameter inference.
Unbiased likelihood estimates for state space models can be produced by \emph{particle filter} algorithms.
Algorithm \ref{alg:BPF} presents the basic \emph{bootstrap particle filter} (BPF) used in this paper, but there are many variations.
For more details see for example \cite{Doucet2011}, \cite{Sarkka2013} and \cite{Fearnhead2018}.
For proof that the particle filter likelihood estimate is indeed unbiased see \cite{DelMoral2004} and \cite{Pitt2010}.

Combining the PMMH algorithm with a particle filter can target $p(\theta | y)$ for state space models.
\cite{Andrieu2010} extend this approach to give \emph{particle MCMC} (pMCMC), which targets the joint posterior $p(\theta, x | y)$; we refer the reader to this paper for a full description of pMCMC.
%DISCUSS THAT WE'RE NOT GOING INTO THIS? RGE: I think a remark referring the reader to the PMCMC paper is fine.

\begin{algorithm}[htb] \caption{Pseudo-marginal Metropolis Hastings} \label{alg:PMMH}
\begin{algorithmic}
\STATE Input: initial state $\theta_0$ and likelihood estimate $\hat{L}_0$, proposal density $q(\theta^*|\theta)$
\FOR{$i=1,2,\ldots$}
\STATE 1. Sample proposal $\theta^*$ from $q(\theta^* | \theta_{i-1})$.
\STATE 2. Calculate $\hat{L}^*$, an estimate of $L(\theta^*)$.
\STATE 3. Accept proposal with probability $\min(1,r)$ where
\[
r = \frac{\hat{L}(\theta^*) \pi(\theta^*) q(\theta_{i-1} | \theta^*)}{\hat{L}(\theta_{i-1}) \pi(\theta_{i-1}) q(\theta^* | \theta_{i-1})}.
\]
Upon acceptance let $\theta_i = \theta^*$ and $\hat{L}_i = \hat{L}^*$.
Otherwise let $\theta_i = \theta_{i-1}$ and $\hat{L}_i = \hat{L}_{i-1}$.
\ENDFOR
\STATE Output: $\theta_1, \theta_2, \ldots$
\end{algorithmic}
\end{algorithm}

\begin{algorithm}[htb] \caption{Bootstrap particle filter. (This algorithm drops $\theta$ from the conditioning for notational simplicity.)} \label{alg:BPF}
\begin{algorithmic}
\STATE Input: number of particles $N$
\STATE \textbf{Initialise}. For $i=1,2,\ldots,N$ sample particle $x_0^{(i)}$ from the initial state distribution and assign weight $w_0^{(i)}=1/N$.
\STATE (Or, if a $y_0$ observation is available, compute weights as in step 3.)
\FOR{$t=1,2,\ldots,T$}
\STATE \textbf{1. Resample}. For $i=1,2,\ldots,N$ sample $\tilde{x}_t^{(i)}$ from the $x_{t-1}^{(j)}$ particles with probabilities $w_{t-1}^{(j)}$. (This step can be omitted for $t=1$ if there is no $y_0$ observation.)
\STATE \textbf{2. Propagate}. For $i=1,2,\ldots,N$ sample $x_t^{(i)}$ from $p(\cdot | \tilde{x}_t^{(i)})$.
\STATE \textbf{3. Weight}. For $i=1,2,\ldots,N$ compute weight $\tilde{w}_t^{(i)} = p(y_t | x_t^{(i)})$
and normalised weight $w_t^{(i)} = \tilde{w}_t^{(i)} / S_t$
where $S_t = \sum_{i=1}^N \tilde{w}_t^{(i)}$.
\ENDFOR
\STATE Output: likelihood estimate $\hat{L} = \prod_{t=1}^T \frac{S_t}{N}$.
\STATE (Or, if a $y_0$ observation is available, take the product from $t=0$.)
\end{algorithmic}
\end{algorithm}

\subsubsection{PMMH tuning} \label{sec:tuning}

PMMH using BPF likelihood estimates has several tuning choices.
This section sets out the approach we use to make these choices in this paper.
Our choices are consistent with the theoretical analyses of \cite{Sherlock2015} and \cite{Doucet2015}, who derive tuning recommendations under two different sets of simplifying assumptions.

We select the number of particles $N$ for the PF prior to running Algorithm \ref{alg:PMMH} so that the estimated log-likelihood at a representative parameter value has a standard deviation of roughly 1.5.  The parameter value used should have good support under the posterior; we typically use marginal posterior medians from exploratory analyses.

We use a normal random walk proposal distribution: $\theta^* \sim \mathcal{N}(\theta_{i-1}, \Sigma)$.
We take $\Sigma$ to be an estimate of the posterior variance, again taken from exploratory analyses.
\cite{Sherlock2015} and \cite{Doucet2015} provide guidance for scaling the variance matrix by a scalar to improve performance, finding that this was helpful for high dimensional target distributions for instance.
We did not find this necessary for our analyses of low dimensional targets, but make use of this approach in Section \ref{subsec:neuroscience} where an 11-dimensional target is considered.

\subsection{Ensemble Kalman Filter} \label{subsec:enkf}

Here, we give a brief overview of the ensemble Kalman filter \citep{evensen1994} and refer the reader to 
\cite{Katzfuss2016} and the references therein for further details.

Consider the task of sampling the filtering density $p(x_t|y_{1:t})$ where $y_{1:t}=(y_1,\ldots,y_t)$.
(We omit explicit conditioning on the parameter vector $\theta$ throughout this section.) The EnKF 
generates approximate draws from $p(x_t|y_{1:t})$ via a sequence of forecasting and updating steps. 
Suppose that a sample $\{x_{t-1}^{(1)},\ldots,x_{t-1}^{(N)}\}$ (known as the \emph{filtering 
ensemble}) is available at time $t-1$ from $p(x_{t-1}|y_{1:t-1})$. The \emph{forecast ensemble} $\{\tilde{x}_t^{(1)},\ldots,\tilde{x}_t^{(N)}\}$ 
is obtained by drawing $\tilde{x}_t^{(i)}\sim p(\cdot|x_{t-1}^{(i)}), i=1,\ldots,N$. 
The forecast density $p(x_t|y_{1:t-1})$ is then approximated by
\[
p_{\textrm{enkf}}(x_t|y_{1:t-1})=\mathcal{N}(x_t\,;\,\hat{\mu}_{t|t-1}\,,\,\hat{\Sigma}_{t|t-1})
\]
where $\mathcal{N}(\cdot;\mu,\Sigma)$ denotes the multivariate Gaussian density with mean $\mu$ and variance matrix 
$\Sigma$. The quantities $\hat{\mu}_{t|t-1}$ and $\hat{\Sigma}_{t|t-1}$ are typically taken to be the sample mean and variance 
computed from the forecast ensemble (some extensions of the EnKF use alternative estimates; see \citet{Katzfuss2016} for some common approaches). Now, given the linear Gaussian form of (\ref{eq:Gaussian observation}), 
the joint distribution of $X_t$ and $Y_t$ (given $y_{1:t-1}$) can be obtained approximately as
\begin{equation}\label{joint}
\begin{pmatrix}X_t\\Y_t\end{pmatrix}\sim N
\left\{
\begin{pmatrix}\hat{\mu}_{t|t-1}\\P\hat{\mu}_{t|t-1}\end{pmatrix}\,,\,
\begin{pmatrix}\hat{\Sigma}_{t|t-1} & \hat{\Sigma}_{t|t-1} P'\\
P\hat{\Sigma}_{t|t-1}  & P\hat{\Sigma}_{t|t-1} P'+S
\end{pmatrix}
\right\}.
\end{equation} 
Hence, conditioning on $Y_t=y_t$ gives 
\begin{equation}\label{enkfPost}
p_{\textrm{enkf}}(x_t|y_{1:t})=\mathcal{N}(x_t\,;\,\hat{\mu}_{t|t}\,,\,\hat{\Sigma}_{t|t})
\end{equation}
where $\hat{\mu}_{t|t}=\hat{\mu}_{t|t-1}+\hat{K}_t(y_t-P\hat{\mu}_{t|t-1})$, $\hat{\Sigma}_{t|t}=(I_{d_x}-\hat{K}_t P)\hat{\Sigma}_{t|t-1}$ 
and $\hat{K}_t$ is an estimate of the Kalman gain, that is
\begin{equation}\label{Kgain}
\hat{K}_t = \hat{\Sigma}_{t|t-1}P'(P\hat{\Sigma}_{t|t-1} P'+S)^{-1}.
\end{equation}
It is then straightforward to generate samples from (\ref{enkfPost}) to be used as the filtering ensemble at the 
next time point. However, rather than explicitly calculate the filtering density in (\ref{enkfPost}), the standard implementation 
of the EnKF \cite[see e.g.][]{Katzfuss2016} performs a \emph{shifting} step, which is equivalent under the Gaussianity assumption \eqref{joint} (and a Gaussian prior for $x_0$). For each particle (known in this context 
as an \emph{ensemble member}), we compute 
$x_{t}^{(i)}=\tilde{x}_t^{(i)}+\hat{K}_t(y_t-\tilde{y}_t^{(i)})$, where $\tilde{y}_t^{(i)}\sim \mathcal{N}(P\tilde{x}_t^{(i)},S)$ 
is a pseudo-observation.
Note that the shifting step only requires a draw from a $d_y$ variate Gaussian distribution per 
particle, rather than a draw from a $d_x$ variate Gaussian if (\ref{enkfPost}) is sampled directly. Moreover, the shifting approach does 
not make the strong assumption that the forecast ensemble is Gaussian distributed. There are also other schemes for performing the shifting that we do not consider here. The randomness in the shifting step leads to the variant of the EnKF described being referred to as the ``stochastic'' EnKF; a commonly used alternative is to use a deterministic shift of the ensemble members \citep{Tippett2003}. In what follows, we typically use the stochastic shifting step.

Given a sample $\{x_0^{(i)},\ldots,x_0^{(N)}\}$ from the state prior, the EnKF recursively 
alternates between computing the forecast ensemble, and shifting each ensemble member, to give approximate draws from 
the filtering density $p(x_t|y_{1:t})$, $t=1,\ldots,T$.
We state a version of the EnKF based on a shifting step below as Algorithm~\ref{alg:EnKF}.

The EnKF is most easily understood in the context of a linear Gaussian state space model. In this special case, 
the filtering distribution $p_{\textrm{enkf}}(x_t|y_{1:t-1})$ converges to the true filtering distribution as the 
number of ensemble members $N\to\infty$. Essentially, the EnKF converges to the Kalman filter. For finite $N$ and 
a linear state space model, the EnKF approximates the Kalman filter by replacing the mean and variance of the 
forecast distribution with their sample equivalents. The resulting dimension reduction (that only requires storing 
and manipulating $d_x$-vectors) avoids the potentially expensive calculation and storage of the forecast variance matrix. 
Moreover, several studies \citep[e.g.][]{Lei2010,Houtekamer2014,Katzfuss2019} have found that the EnKF shifting step works well for non-Gaussian evolution densities. 
We therefore consider the use of the EnKF likelihood inside a Metropolis-Hastings scheme. We provide a motivation 
and give details of our proposed approach in the next section.

\section{Ensemble MCMC} \label{sec:method}

It is well known that as the variance of the likelihood estimator increases, the acceptance probability of the pseudo-marginal MH scheme rapidly decreases to 0 \citep{pitt12}, 
resulting in slow mixing behaviour of the parameter chains. As discussed in Section~\ref{sec:tuning}, 
a value of $N$ (the number of particles) can be chosen to balance mixing performance and computational cost. 
Nevertheless, in scenarios where the stochasticity inherent in the state process dominates the 
observation variance, the number of particles required to maintain a reasonable likelihood variance is 
likely to render BPF-driven PMMH computationally infeasible. Methods that aim to alleviate this 
problem include the use of an auxiliary particle filter \citep[see e.g.][]{GoliWilk15}, 
%and correlated PMMH \citep{deligiannidis2016,dahlin2015}. The former requires careful exploitation of the 
%model structure in order to propagate particles conditional on the observations, whereas the latter 
%requires additional sorting steps prior to resampling, and the specification of an additional tuning parameter 
%that controls the correlation between successive likelihood estimates. 
which requires careful exploitation of the model structure in order to propagate particles conditional on the observations. 
Our proposed approach is simple to implement and, for the simplest implementation, does not require the specification of any additional tuning parameters. 

Here we outline our proposed \emph{ensemble MCMC} (eMCMC) algorithm.
In essence, this is PMMH using the EnKF as fast replacement for the BPF to estimate the likelihood $L(\theta)$.
We describe the eMCMC algorithm in more detail now, and in the following subsections we discuss some extensions to improve its efficiency.

First we derive a likelihood estimate based on EnKF calculations.
Recall that the (marginal) likelihood can be factorised as
\begin{equation}\label{like}
L(\theta) = p(y_1|\theta)\prod_{t=2}^{T}p(y_t|y_{1:t-1},\theta).
\end{equation}
From \eqref{joint} it follows that an EnKF approximation of $p(y_t|y_{1:t-1},\theta)$ is 
\[
p_{\textrm{enkf}}^N(y_t|y_{1:t-1},\theta)=\mathcal{N}(y_t\,;\,P\hat{\mu}_{t|t-1}\,,\, P\hat{\Sigma}_{t|t-1} P'+S)
\]
which can easily be computed for each $t=1,\ldots,T$, with the notational convention that $p(y_1|\theta)=p(y_t|y_{1:0},\theta)$.  The need for the explicit dependence on $N$ for this likelihood estimate will become clearer later in this section.  The overall approximation to the likelihood is given by
\begin{equation} \label{enkf like}
L_{\textrm{enkf}}^N(\theta) = \prod_{t=1}^{T}p_{\textrm{enkf}}^N(y_t|y_{1:t-1},\theta).
\end{equation}
The EnKF including likelihood estimation is given by Algorithm~\ref{alg:EnKF}.
The ensemble MCMC scheme is then implemented by running Algorithm~\ref{alg:PMMH} with $\hat{L}$ replaced by $\hat{L}_{\textrm{enkf}}^N$.
One issue in implementing eMCMC is how to perform tuning.
Due to the absence of specialised theory, we use the same tuning guidance as for PMMH with BPF likelihood estimates, described above in Section \ref{sec:tuning}.

\begin{algorithm}[htb] \caption{Ensemble Kalman filter} \label{alg:EnKF}
\begin{algorithmic}
\STATE Input: number of ensemble members $N$
\STATE \textbf{Initialise.} For $i=1,2,\ldots,N$ sample $x_0^{(i)}$ from the initial state distribution. Set $\hat{L}_{\textrm{enkf}}=1$.
\FOR{$t=1,2,\ldots,T$}
\STATE \textbf{1. Forecast ensemble}. For $i=1,2,\ldots,N$ sample $\tilde{x}_t^{(i)}\sim p(\cdot|x_{t-1}^{(i)})$.
\STATE \textbf{2. Likelihood update}. Compute estimates of the forecast mean and variance: $\hat{\mu}_{t|t-1}$ and $\hat{\Sigma}_{t|t-1}$. Set $\hat{L}_{\textrm{enkf}}:=\hat{L}_{\textrm{enkf}}\times \mathcal{N}(y_t\,;\,P\hat{\mu}_{t|t-1}\,,\, P\hat{\Sigma}_{t|t-1} P'+S)$.
\STATE \textbf{3. Shift ensemble}. Compute the approximate Kalman gain, $\hat{K}_t$, given by (\ref{Kgain}). For $i=1,2,\ldots,N$, set $x_{t}^{(i)}=\tilde{x}_t^{(i)}+\hat{K}_t(y_t-\tilde{y}_t^{(i)})$, where $\tilde{y}_t^{(i)}\sim \mathcal{N}(P\tilde{x}_t^{(i)},S)$ is a pseudo-observation.
\ENDFOR
\STATE Output: likelihood estimate $\hat{L}_{\textrm{enkf}}$
\end{algorithmic}
\end{algorithm}

It is worth emphasising that, unlike pMCMC, the eMCMC posterior
\begin{align*}
p_{\textrm{enkf}}^N(\theta|y) \propto L_{\textrm{enkf}}^N(\theta) p(\theta),
\end{align*}
does not in general equal the posterior $\pi(\theta|y)$ exactly. The reason is that, unlike the BPF, the EnKF gives a biased estimator of $L(\theta)$, precluding its use for exact approximate inference. Nevertheless, as noted by \cite{Stroud2010}, \cite{Stroud2018} 
and \cite{Katzfuss2019}  among others, the variance of the likelihood estimator under the EnKF can be relatively small, suggesting that 
use of EnKF inside a Metropolis-Hastings scheme is likely to be of practical use, particularly in scenarios when the BPF is computationally prohibitive. 

In fact, even when the forecast ensemble is exactly Gaussian distributed for all $t$, the EnKF posterior still does not target the exact posterior, since $\mathcal{N}(y_t\,;\,P\hat{\mu}_{t|t-1}\,,\, P\hat{\Sigma}_{t|t-1} P'+S)$ is a biased estimate of the \emph{idealised} normal density should we be able to take $N \rightarrow \infty$.  Thus, for finite $N$, the eMCMC target is not the idealised eMCMC target, $p_{\textrm{enkf}}^\infty(\theta|y)$.  However, we find empirically that our method appears to be weakly dependent on $N$.  Given this, we suggest to choose $N$ to maximise the computational efficiency by borrowing similar advice from the pseudo-marginal literature (as described in Section \ref{sec:tuning}).  Interestingly, there is an exactly unbiased estimator of a normal density given a sample from it, and we exploit this in Section \ref{subsec:unbiased}.  We discuss the unbiased version and other extensions below.  

\subsection{Randomised Quasi-Monte Carlo} \label{subsec:rqmc}

For this subsection, all quantities are conditioned on $\theta$ so we drop it for notational convenience. At iteration $t$ of the EnKF we are interested in estimating $\mu_{t|t-1}$ and $\Sigma_{t|t-1}$, so that we can approximate the conditional likelihood $p(y_t|y_{1:t-1})$ with a Gaussian density.  These moments can be estimated via firstly performing the shifting step at $t-1$ and conditional on the result simulating from the forward evolution density.  This is effectively an approximate sample from the joint distribution $p(x_t|x_{t-1})p(x_{t-1}|y_{1:{t-1}})$.  Then $\mu_{t|t-1}$ and $\Sigma_{t|t-1}$ are estimated from the $N$ ensemble members.  

Often it is possible to write the simulation from a standard statistical distribution as a function of a uniform random number.   For example, to simulate from a $y \sim \mathcal{N}(\mu,\sigma^2)$ distribution we can compute the following, $y = \mu + \sigma\cdot \Phi^{-1}(u)$ where $u \sim \mathcal{U}(0,1)$ and $\Phi^{-1}(u)$ is the quantile function of the standard normal density.  Assume that we can write the evolution density as a function of $m$ uniform random variates.  Then, we require $d_y + m$ uniform random numbers to approximately simulate from $x_t,x_{t-1}|y_{1:{t-1}}$ ($d_y$ for the shifting step and $m$ for simulating the evolution density).  Given $N$ particles, we use $N \times (d_y + m)$ uniform random numbers for estimating $\mu_{t|t-1}$ and $\Sigma_{t|t-1}$.  The naive approach is to draw these via pseudo-random numbers.  However, significant variance reduction could be achieved by simulating from the $(d_y + m)$-dimensional object $N$ times using \emph{randomised quasi-Monte Carlo} (RQMC, \citealp{Niederreiter:1992}).  QMC is well known to generate a sequence of numbers that have superior space filling properties in the unit hypercube compared to pseudo-random numbers.  The randomised component ensures that expectations can be estimated unbiasedly.  Random numbers from the joint distribution of interest, $x_t,x_{t-1}|y_{1:{t-1}}$, can be achieved via transforming the RQMC numbers as recently discussed.  We use this approach to bring down the variance of the estimators of $\mu_{t|t-1}$ and $\Sigma_{t|t-1}$, which hopefully reduces the variance of the estimator for $p(y_t|y_{1:t-1})$.  For generating the RQMC numbers in this paper, we use the scrambled Sobol's net, i.e.~the scrambled $(t,m,s)$-net in base $b=2$.

RQMC has recently received increasing attention in the statistics community.  \cite{Tran:2015} document a faster convergence in their Variational Bayes updating procedure
when the noisy gradient is computed using RQMC, \citet{Drovandi2018} use it to reduce the variance of expected utility estimation within Bayesian optimal design and
\cite{Gerber2015} show the efficiency of RQMC in particle filtering.
However their application to particle filtering requires considerable ingenuity (using a Hilbert curve method to perform resampling). It is interesting to note the ease with which RQMC can be exploited in the EnKF in comparison.

\subsection{Correlated eMCMC} \label{subsec:correlated}

As mentioned above, the EnKF requires generating random numbers for the shifting step and simulating the evolution density.  The former can be generated by standard normal random variates and the Cholesky factorisation of the covariance matrix.  We assume in this section that the evolution density can be simulated either directly or indirectly via a suitable transformation with standard normal random numbers.  Denote the collection of these random numbers required in the EnKF as $u$.

\citet{deligiannidis2018correlated} and \citet{dahlin2015accelerating} develop the \emph{correlated pseudo-marginal MCMC} method where they consider the joint target density $p(\theta,u|y)$ where $u$ are random numbers required to estimate the likelihood unbiasedly, $p(y|\theta,u)$.  It is easy to show that the $\theta$-marginal of the joint distribution is the posterior of interest, $p(\theta|y)$.  Assume that $u$ are independent standard normal random variates.  The idea of correlated pseudo-marginal is to induce correlation in successive likelihood estimates in MCMC by correlating the $u$ random numbers. This can have the effect of mitigating ``sticky'' behaviour often seen in pseudo-marginal chains since, in the correlated scheme, if the likelihood is overestimated at the current iteration, it is also likely to be overestimated at the next.  The joint proposal distribution of the correlated pseudo-marginal method is given by $q(\theta^*,u^*|\theta,u) = q(\theta^*|\theta)\mathcal{N}(u^*;\sqrt{1-\sigma_u^2}u, \sigma_u^2 \mathcal{I})$, where $\mathcal{I}$ is the identity matrix.  The proposal for $u$ is the \emph{Crank-Nicolson proposal} and it is invariant with respect to the marginal distribution of $u$.  $\sigma_u^2$ is an additional tuning parameter that is typically set to be small so that $u^*$ is highly correlated with $u$.

Here we consider applying this correlated pseudo-marginal approach to our eMCMC method, with the motivation that a smaller ensemble size $N$ can be used, reducing computational cost. 
Note that BPF driven pMCMC requires additional modification to accommodate this approach, as it did for RQMC. Essentially, the resampling step has the effect of breaking down correlation between successive likelihood estimates. To alleviate this problem, the particles can be sorted before propagation e.g.\ using a Hilbert sorting procedure \citep{deligiannidis2018correlated} or simple Euclidean sorting \citep{choppala2016}. The random numbers used in the resampling step itself should also be updated using the Crank-Nicolson proposal. Since the eMCMC scheme does not use resampling, incorporating correlation is straightforward.

\subsection{Unbiased Ensemble Kalman Filter Likelihood} \label{subsec:unbiased}

As mentioned earlier, even if the sample from the forecast distribution was exactly Gaussian for some $t$, the corresponding EnKF likelihood estimate for $y_t$ would not be unbiased.   In general, for some data $y$ and a sample of size $N$ from a Gaussian distribution, $x = x_1,\ldots,x_N \sim \mathcal{N}(\mu,\sigma)$, the density estimator $\mathcal{N}(y;\mu_N,\Sigma_N)$ is not an unbiased estimator of $\mathcal{N}(y;\mu,\Sigma)$ where $\mu_N$ and $\Sigma_N$ are the sample mean covariance computed from the sample $x$.  Given the bias present in the EnKF likelihood estimate, even when the Gaussian assumption is correct, the EnKF posterior, unlike standard pseudo-marginal, theoretically depends on $N$.

Even though we demonstrate empirically in Section \ref{sec:results} that the eMCMC posterior seems to be only weakly dependent on $N$, we present a new approach now that will likely be less sensitive to $N$.  Interestingly, there does exist an unbiased estimator of a Gaussian density given only an iid sample from the same Gaussian density.   Using the notation of \citet{Ghurye1969}, let
\begin{align*}
c(k,v) &= \frac{2^{-kv/2}\pi^{-k(k-1)/4}}{\prod_{i=1}^k\Gamma \left(\frac{1}{2}(v-i+1)\right)},
\end{align*}
and for a square matrix $A$ write $\psi(A) = |A|$ if $A>0$ and $\psi(A) = 0$ otherwise, where $|A|$ is the determinant of $A$ and $A > 0$ means that $A$ is positive definite.  The result of \citet{Ghurye1969} shows that an exactly unbiased estimator of ${\cal N}(y;\mu,\Sigma)$ is (in the case where $y$ is Gaussian and $N > d+3$ where $d$ is the dimension of $y$)
\begin{align*}
\widehat{\mathcal{N}}(y;\mu,\Sigma) &= (2\pi)^{-d/2}\frac{c(d,N-2)}{c(d,N-1)(1-1/N)^{d/2}}|M_N|^{-(N-d-2)/2}  \\ 
& \qquad \psi \left(  M_N - (y - \mu_N)(y - \mu_N)^\top/(1-1/N)   \right)^{(N-d-3)/2},
\end{align*}
where $M_N = (N-1)\Sigma_N$.  We propose to replace the standard Gaussian density estimator in the EnKF likelihood estimator with this alternative estimator.  Note that this estimator has also been used in \citet{price+dln16} for approximating intractable likelihoods in simulation-based likelihood-free estimation problems. 

We refer to the method when we use the unbiased Gaussian density estimator in the EnKF likelihood estimator as ueMCMC.  We stress that this approach still does not target the true posterior, but at least will not depend on the number of ensemble members $N$ when the Gaussian assumption is correct, i.e.\ the target is exactly the idealised approximation, $p_{\textrm{enkf}}^\infty(\theta|y)$.  Even though the forecast density is unlikely to be exactly Gaussian in practice, we do expect ueMCMC to be less sensitive to $N$ compared with eMCMC.  We note that this method might be particularly useful when combined with the correlated approach in Section \ref{subsec:correlated}, since it might be sufficient to use a very small $N$ to achieve reasonable computational efficiency, but the small $N$ may produce bias in the eMCMC posterior compared to the idealised eMCMC posterior.

\subsection{Early rejection} \label{subsec:rejection}

\citet{Prangle2018} apply pMCMC in the setting of approximate Bayesian computation (ABC). They outline a method for rejecting proposed values of $\theta$ that
have a small estimated likelihood without running the whole particle
filter. In \citet{Everitt2019} it is shown that this approach
can be extended to pMCMC when using the BPF. A similar approach
may be used in eMCMC. Suppose that the likelihood
estimate from the EnKF is implemented sequentially, as the EnKF is
running. Recall from \eqref{enkf like} that the EnKF likelihood estimate is a product,
$L_{\textrm{enkf}}^N(\theta) = \prod_{t=1}^{T} \alpha_t$.
Here $\alpha_t = \mathcal{N}\left(y_{t}; P\hat{\mu}_{t\mid t-1},P\Sigma_{t\mid t-1}P'+S\right)$ is calculated in iteration $t$ of the EnKF.
We are guaranteed that an upper bound on $\alpha_t$ is given by $B(\theta^*):=\mathcal{N}\left(\mathbf{0};\mathbf{0},S(\theta^*) \right)$.
Thus we have that $\alpha_t/B \leq 1$.
This fact ensures that %$L_{\textrm{enkf}}^N(\theta) \leq B \left(\theta^{*}\right)^T$ and that
$\hat{r}_{\text{enkf}}^{(\tau)} := \prod_{t=1}^\tau \alpha_t/B$ is an upper bound on $L_{\textrm{enkf}}^N(\theta) / B^T$ which can be calculated at iteration $\tau$ of the EnKF.

We use this property to propose an ``early rejection'' algorithm.
The idea is that during an EnKF run, as soon as $\hat{r}_{\text{enkf}}$ drops below a certain threshold, we are sure that the MCMC proposal $\theta^{*}$ will not be accepted.
Hence we can save time by immediately terminating the EnKF run.
Algorithm \ref{alg:early_rejection} describes a single iteration of the resultant MCMC algorithm, which involves reorganising the order of calculation of the acceptance probability and likelihood estimate from our standard eMCMC algorithm.
This early rejection approach is employed
in Section \ref{subsec:neuroscience}, where a computationally
expensive model is studied.

\begin{algorithm}[htb] \caption{An iteration of early-rejection eMCMC.} \label{alg:early_rejection}
\begin{algorithmic}
\STATE \textbf{Input:} $\theta$, the current value of the parameter and $L_{\text{enkf }}^{N}(\theta)$, the estimate of the likelihood for this parameter.
\STATE Simulate $\theta^{*}\sim q\left(\cdot\mid\theta\right)$, and let
$S\left(\theta^{*}\right)$ be the measurement noise matrix for this
proposed parameter.
\STATE Let $B\left(\theta^{*}\right)=\mathcal{N}\left(\mathbf{0};\mathbf{0},S\left(\theta^{*}\right)\right)$.
\STATE Simulate $u\sim\mathcal{U}\left(0,1\right)$.
\STATE \textbf{Initial EnKF step:} for $i=1 \ldots N$, simulate $x_{0}^{(i)}$ from the initial state distribution.
\STATE Initialise estimate $\hat{r}_{\text{enkf}}=1$, then perform first \textbf{early rejection} step:
\IF{$\hat{r}_{\text{enkf}}<u\frac{p\left(\theta\right)L_{\text{enkf }}^{N}\left(\theta\right)}{p\left(\theta^{*}\right)}\frac{q\left(\theta^{*}\mid\theta\right)}{q\left(\theta\mid\theta^{*}\right)}\frac{1}{B^T \left(\theta^{*}\right)}$}
\STATE reject $\theta^{*}$ and break.
\ENDIF
\FOR{$t=1 \ldots T$}
\STATE \textbf{1. Forecast ensemble}. For $i=1,\ldots,N$ sample $\tilde{x}_{t}^{(i)}\sim p(\cdot|x_{t-1}^{(i)})$.
\STATE \textbf{2. Likelihood update}. Compute estimates of the forecast mean
and variance: $\hat{\mu}_{t|t-1}$ and $\hat{\Sigma}_{t|t-1}$. Set
$\hat{r}_{\textrm{enkf}}:=\hat{r}_{\textrm{enkf}}\times \mathcal{N}\left(y_{t}\,;\,P\hat{\mu}_{t|t-1}\,,\,P\hat{\Sigma}_{t|t-1}P'+S\left(\theta^{*}\right)\right)/B\left(\theta^{*}\right)$.
\STATE \textbf{3. Early rejection}.
\IF{$\hat{r}_{\text{enkf}}<u\frac{p\left(\theta\right)L_{\text{enkf }}^{N}\left(\theta\right)}{p\left(\theta^{*}\right)}\frac{q\left(\theta^{*}\mid\theta\right)}{q\left(\theta\mid\theta^{*}\right)}\frac{1}{B^T \left(\theta^{*}\right)}$}
\STATE reject $\theta^{*}$ and break.
\ENDIF
\STATE \textbf{4. Shift ensemble}. Compute the approximate Kalman gain, $\hat{K}_{t}$,
given by \eqref{Kgain}. For $i=1,\ldots,N$, set
$x_{t}^{(i)}=\tilde{x}_{t}^{(i)}+\hat{K}_{t}(y_{t}-\tilde{y}_{t}^{(i)})$,
where $\tilde{y}_{t}^{(i)}\sim \mathcal{N}\left(P\tilde{x}_{t}^{(i)},S\left(\theta^{*}\right)\right)$
is a pseudo-observation.
\ENDFOR
\STATE Accept $\theta^{*}$ and let $L_{\text{enkf }}^{N}\left(\theta^{*}\right)=\hat{r}_{\textrm{enkf}}B^T \left(\theta^{*}\right)$.
\end{algorithmic}
\end{algorithm}

\section{Results} \label{sec:results}

Here we demonstrate the potential of our method on several examples with different kinds of complexity.  We select the number of particles $N$ and MCMC proposal variance as described in Section \ref{sec:tuning}.  Given that the different methods have different target distributions, we tune the random walk covariance matrix individually for each method.   

In terms of accuracy we compare the approximate eMCMC and the `exact' pMCMC approach visually.  We note that in many applications it might not be critical to obtain samples from the exact posterior given the potential for model misspecification and/or high accuracy not being important for the analysis aims.  When we deem the eMCMC approximation to be reasonable enough, we compare the statistical efficiency of the two methods using the multivariate effective sample size (ESS) of \citet{Vats2019}.  The overall efficiency considers the statistical efficiency and computing time simultaneously.

In Section \ref{sec:discussion} we provide suggestions on how `exact' posterior sampling can be achieved whilst still using the EnKF.  However, the statistical efficiency gains of these approaches will be reduced compared to eMCMC.

\subsection{Population Ecology Example} \label{subsec:ecology}

\subsubsection{Model and inference task}

\citet{Peters2010} consider a set of competing non-linear state-space population models in ecology and apply them to several datasets.  Denoting the observation at time $t$ as $y_t$ and the corresponding hidden state as $n_t$, the four models we consider are defined below:  
\begin{enumerate}
	\item Ricker model: $\log n_{t+1} = \log n_t + \beta_0 + \beta_1 n_t + \epsilon_t.$
	\item Theta-logistic model: $\log n_{t+1} = \log n_t + \beta_0 + \beta_2 n_t^{\beta_3} + \epsilon_t.$
	\item Mate-limited model: $\log n_{t+1} = 2\log n_t + \beta_0 + \beta_1 n_t - \log(\beta_4 + n_t) + \epsilon_t.$
	\item Flexible-Allee model: $\log n_{t+1} = \log n_t + \beta_0 + \beta_1 n_t + \beta_5 n_t^2 +  \epsilon_t.$
\end{enumerate}
Here $\epsilon_t \sim \mathcal{N}(0,\sigma_w^2)$.  The observation process is assumed to be Gaussian, $y_t|n_t \sim \mathcal{N}(\log n_t, \sigma_e^2)$.  See \citet{Peters2010} for a justification and some qualitative analyses of these models.  The parameters are assumed independent \emph{a priori} and have the following specifications: $\beta_0, \beta_1, \beta_3, \beta_5 \sim \mathcal{N}(0,1)$, $\beta_4, \sigma_w, \sigma_e \sim \mathcal{E}\!xp(1)$ and $\log n_0$ has an improper uniform prior over the real line.

\begin{figure}[tbhp]
	\begin{center}
		\includegraphics[width=0.5\textwidth]{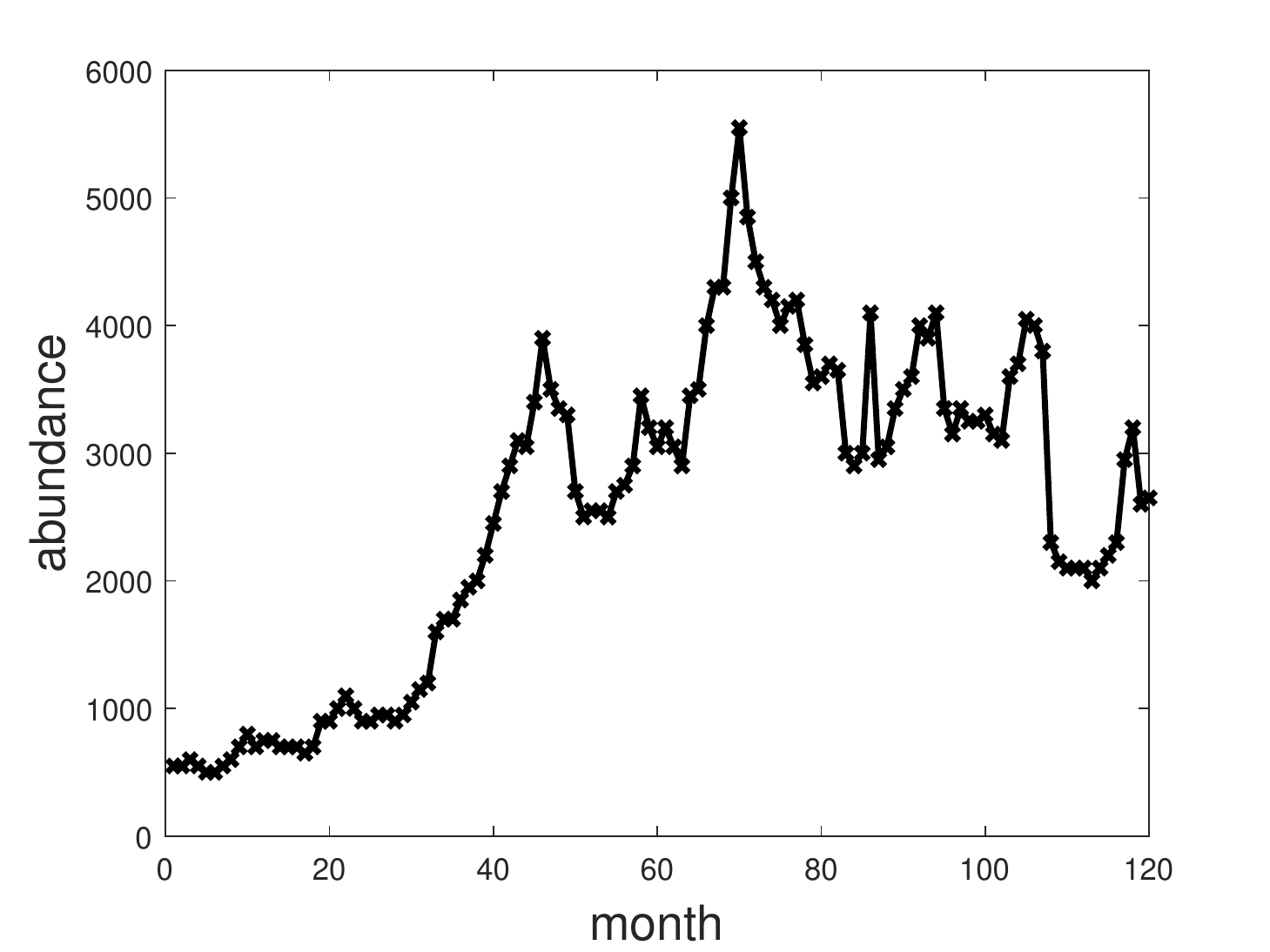}
	\end{center}
	\caption{The nutria dataset.  The observations are shown as crosses and the solid line is a linear interpolation between observations.}
	\label{fig:data_nutria}
\end{figure}

Here we re-analyse the nutria dataset, a time series of female nutria abundance in East Anglia at monthly intervals, considered in \citet{Peters2010} and some references therein.  The data is shown in Figure \ref{fig:data_nutria}. 

\subsubsection{Inference}

It is likely that all the considered models are misspecified but we would like a robust method for fitting them in order to compare the models and investigate possibilities for extending the models.  We find that all models have particular difficulty in capturing the sudden drop in abundance between months 107 and 108.  Further, there appears to be only small observation error.  The consequence for the bootstrap filter is a very small ESS and high variance estimates of the likelihood unless a very large number of particles is used.

For eMCMC, we only require $N=250$ (Ricker, Flexible-Allee, theta-logistic) and $N=200$ (mate-limited) particles.  In contrast, we use $N=50000$ for pMCMC.   For some of the models, the standard deviation of the estimated log-likelihood is still larger than 1.5 even with this large number of particles.  However, we find that when these occur the distribution of the log-likelihood estimator with the BPF has a skew-left distribution, which is less problematic for pMCMC getting stuck at overestimated log-likelihood values.  We find that the pMCMC acceptance rates remain reasonable with $N=50000$ particles.

\begin{figure}[tbhp]
	\begin{center}
		\includegraphics[width=0.7\textwidth]{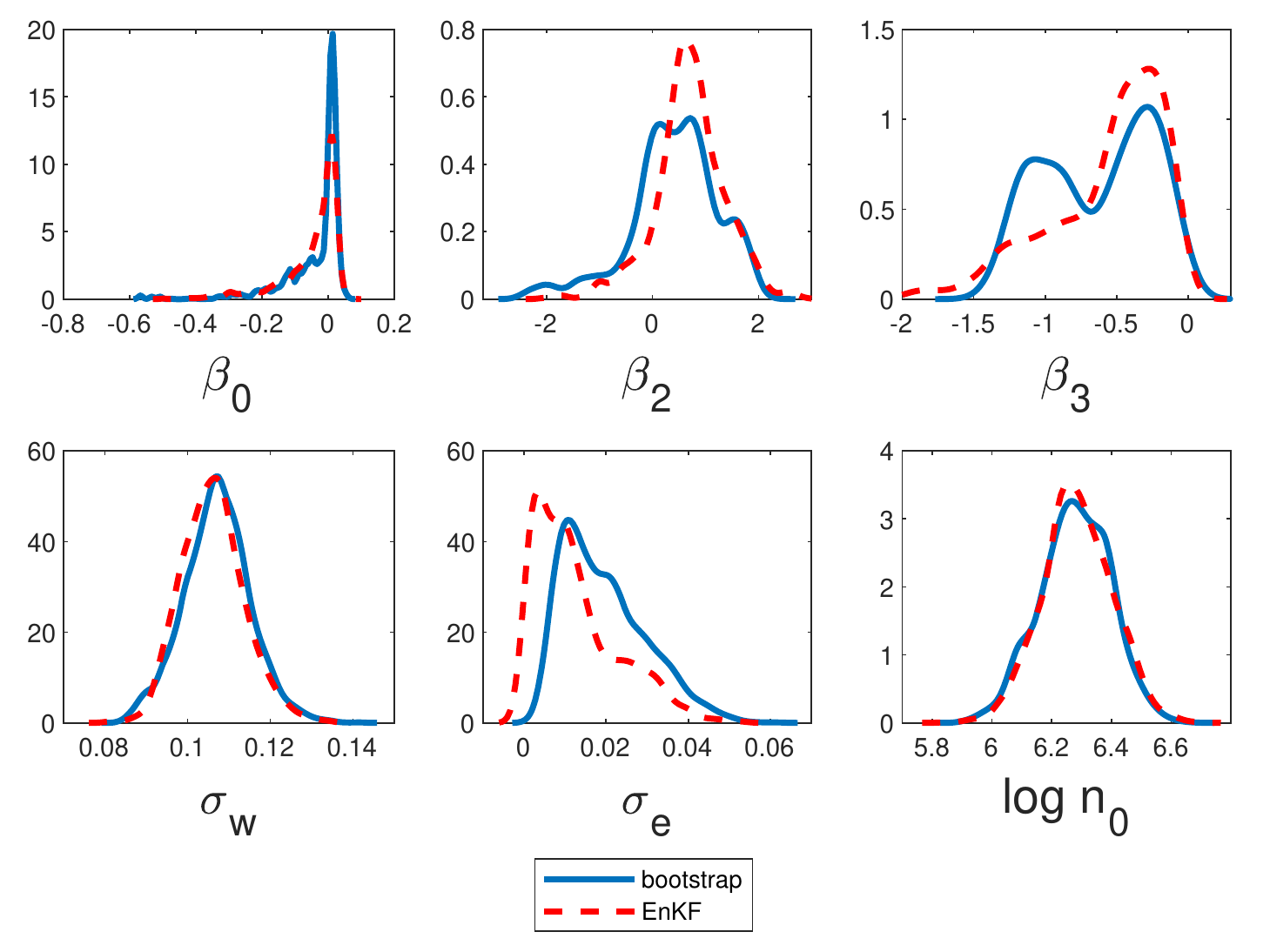}
	\end{center}
	\caption{Estimated univariate posterior distributions for the parameters of the theta-logistic model based on pMCMC (blue solid) and eMCMC (red dash).  }
	\label{fig:ecol_posteriors_thetalog}
\end{figure}

The MCMC acceptance rates for the four models are 15\%, 4\%, 11\% and 10\% (eMCMC), and 8\%, 3\%, 6\% and 5\% (pMCMC), respectively.  The acceptance rates are lower for the theta-logistic model as the posterior distribution is far more irregular compared to the other three models (see Figure \ref{fig:ecol_posteriors_thetalog}).

\begin{figure}[tbhp]
	\begin{center}
		\includegraphics[width=0.7\textwidth]{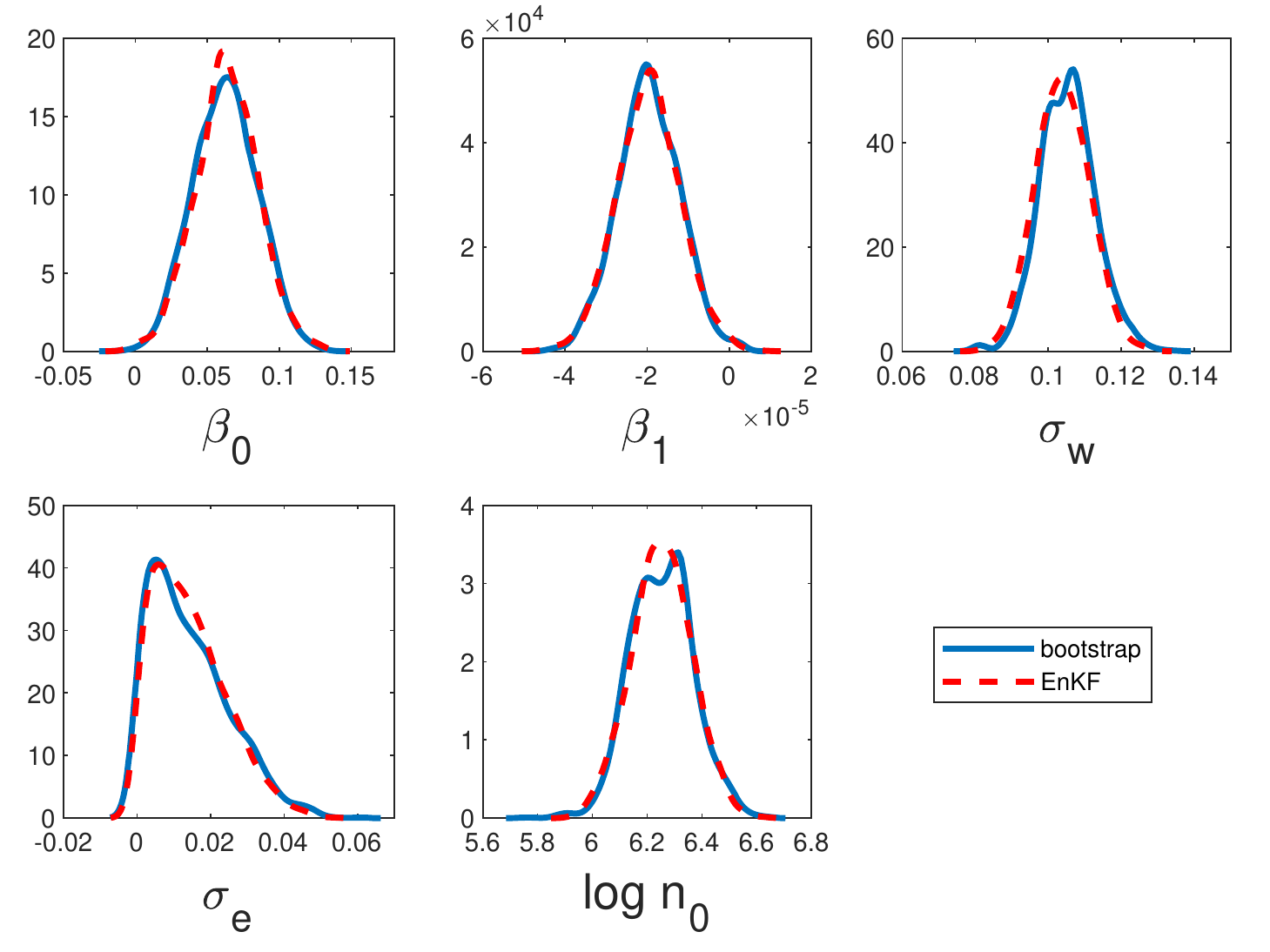}
	\end{center}
	\caption{Estimated univariate posterior distributions for the parameters of the Ricker model based on pMCMC (blue solid) and eMCMC (red dash).  }
	\label{fig:ecol_posteriors_ricker}
\end{figure}

\begin{figure}[tbhp]
	\begin{center}
		\includegraphics[width=0.7\textwidth]{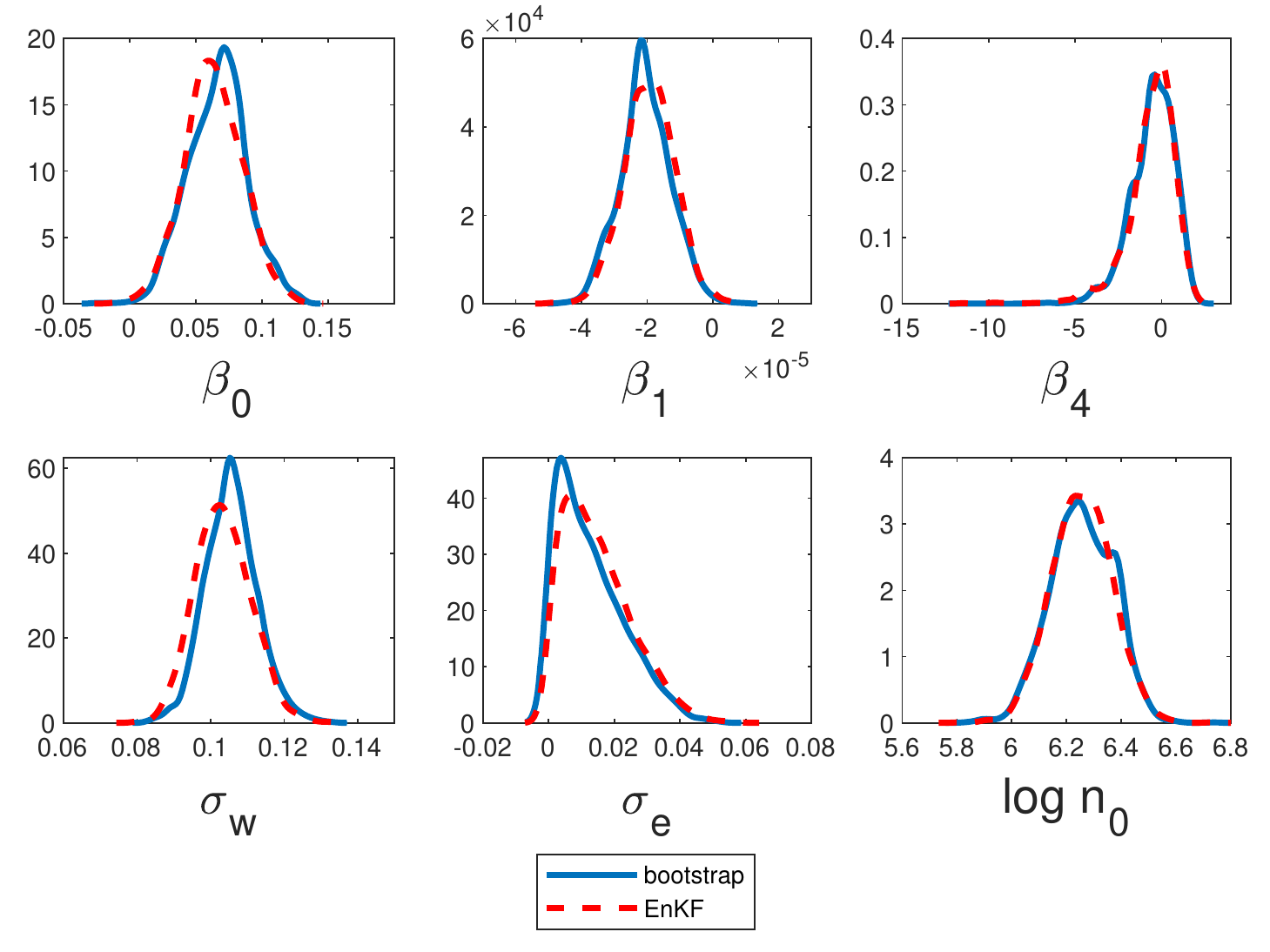}
	\end{center}
	\caption{Estimated univariate posterior distributions for the parameters of the mate-limited model based on pMCMC (blue solid) and eMCMC (red dash).  }
	\label{fig:ecol_posteriors_matelimited}
\end{figure}

\begin{figure}[tbhp]
	\begin{center}
		\includegraphics[width=0.7\textwidth]{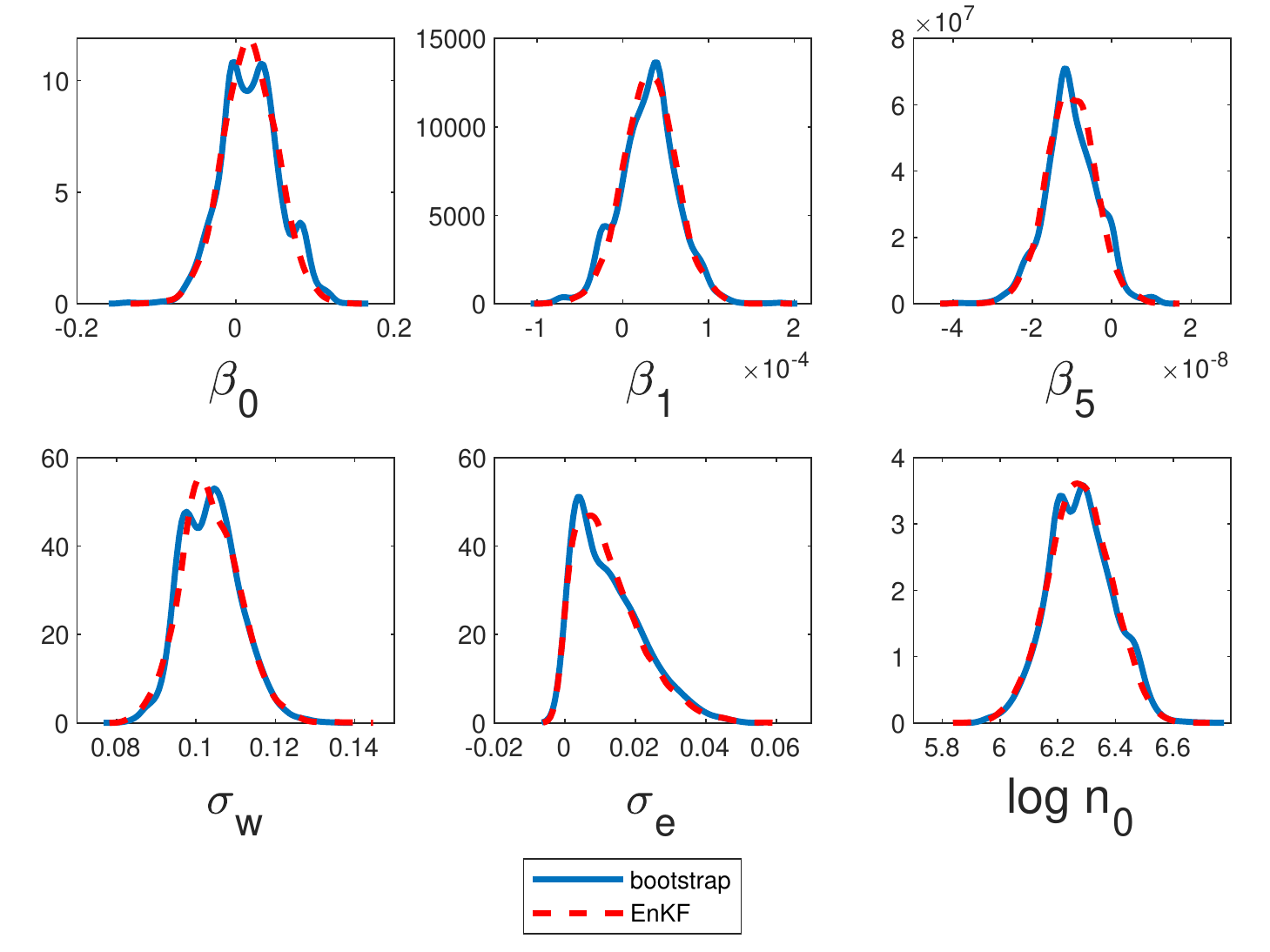}
	\end{center}
	\caption{Estimated univariate posterior distributions for the parameters of the flexible-allee model based on pMCMC (blue solid) and eMCMC (red dash).  }
	\label{fig:ecol_posteriors_flexricker}
\end{figure}

Based on Figures \ref{fig:ecol_posteriors_ricker}, \ref{fig:ecol_posteriors_matelimited} and \ref{fig:ecol_posteriors_flexricker} eMCMC obtains estimated univariate posterior distributions that are remarkably similar to pMCMC.  There is more difference for the theta-logistic model (Figure \ref{fig:ecol_posteriors_thetalog}) but they remain broadly similar.  Further, the Monte Carlo error is greater for this model, potentially exaggerating the differences.

The two methods are compared in terms of computational efficiency on the four models in Table \ref{tab:ecol_efficiency}.  It is evident that the eMCMC approach is producing a two order of magnitude improvement in terms of computational efficiency and still produces reasonable approximations of the posterior.

\begin{table}
	\centering
	\begin{tabular}{cccccc}
		\hline
		Model & filter & $N$ & ESS  & Time (h) & ESS/Time \\
		\hline
		Ricker & BPF & 50000 &  920 & 36.8 & 25 \\
		Ricker & EnKF & 250 & 2400 & 0.14 & 17000  \\  
		Ricker & EnKF + correlation & 25 & 2100  & 0.07  & 30000   \\
		\hline
		theta-logistic & BPF & 50000 & 410 & 40.8 & 10 \\
		theta-logistic & EnKF & 250 & 500 & 0.48 & 1040 \\  
		theta-logistic & EnKF + correlation & 25 & 540  & 0.06  & 9000  \\ 
		\hline
		mate-limited & BPF & 50000 & 770 & 37.6 & 20 \\
		mated-limited & EnKF & 200 & 1460 & 0.35  & 4200 \\  
		mated-limited & EnKF + correlation & 25 & 1800 & 0.07  & 25700 \\  
		\hline
		flexible-allee & BPF & 50000 & 750 & 37.0 & 20 \\
		flexible-allee & EnKF & 250 & 1750 & 0.26  & 6700 \\  
		flexible-allee & EnKF + correlation & 25 & 1600  & 0.08   & 20000  \\  
		\hline
	\end{tabular}
	\caption{Efficiency comparisons for the four non-linear population ecology models.}
	\label{tab:ecol_efficiency}
\end{table}

We test eMCMC for a range of $N$ values in 100-1000 and find the univariate posteriors to show little sensitivity to $N$ (results not shown).  For $N=100$, the MCMC acceptance rate drops substantially and reducing $N$ further is likely to significantly reduce the statistical efficiency of MCMC due to the high-variance likelihood estimates.  Therefore, it is difficult to test the sensitivity of the results to small $N$.  

\begin{figure}[tbhp]
	\begin{center}
		\includegraphics[width=0.7\textwidth]{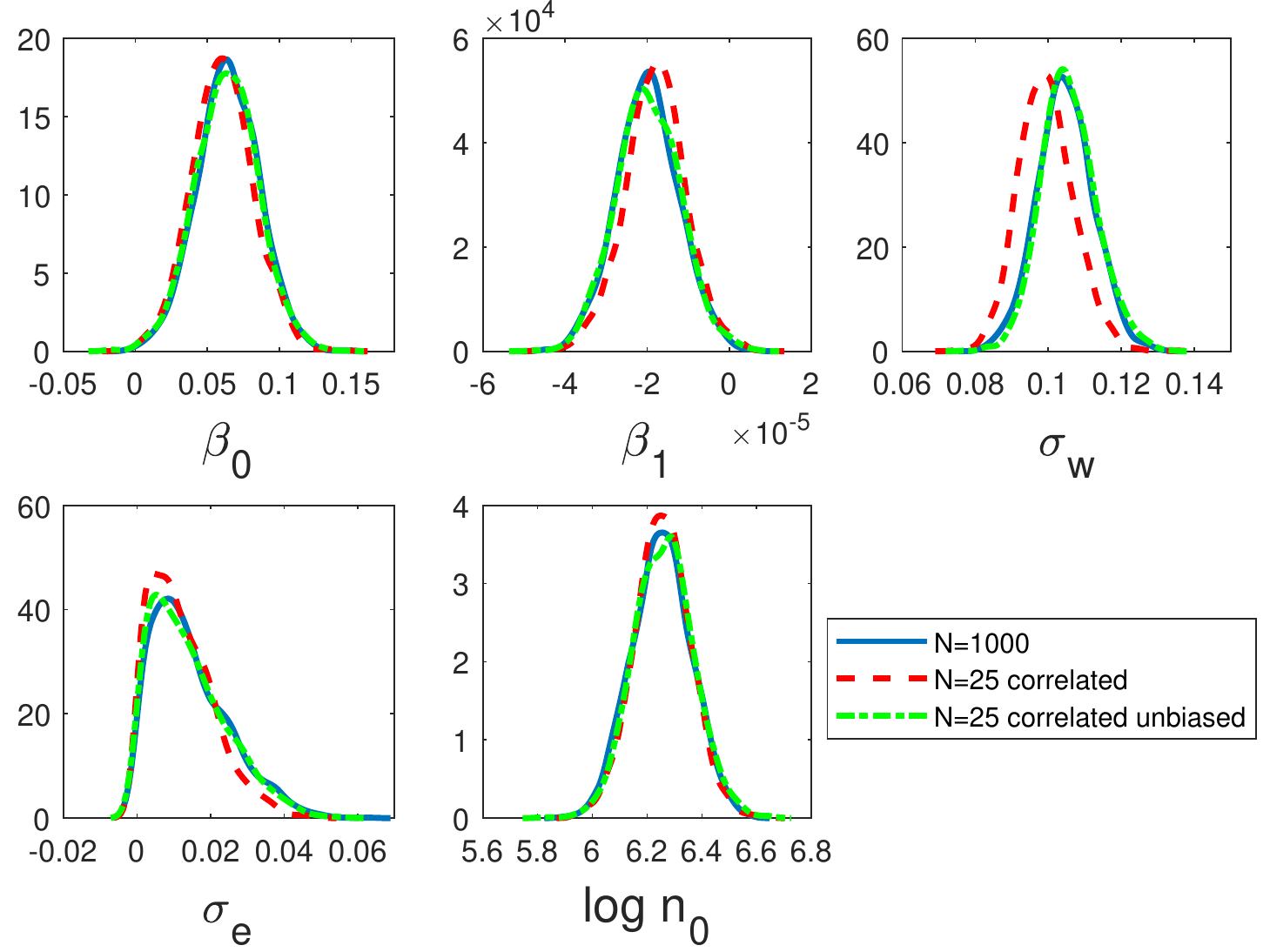}
	\end{center}
	\caption{Approximate univariate posteriors for the Ricker model using eMCMC with $N=1000$ (blue solid), correlated eMCMC with $N=25$ (red dash) and correlated ueMCMC with $N=25$ (green dash-dot).}
	\label{fig:ecol_posteriors_ricker_correlated}
\end{figure}

However, using the correlated extension (with $\sigma_u = 0.1$) allows us to use small $N$ and maintain statistically efficient results.  Similar MCMC acceptance rates as eMCMC with $250$ particles can be achieved using only $N=25$ particles.  The efficiency results can be seen in Table \ref{tab:ecol_efficiency}.  It is evident that the correlation further improves the computational efficiency in this example.  The resulting approximate marginal posteriors compared to eMCMC with $N=1000$ are shown in Figures \ref{fig:ecol_posteriors_ricker_correlated}-\ref{fig:ecol_posteriors_flexricker_correlated} for the four models in Appendix A.  It is evident that similar approximate posteriors are obtained even with vastly different $N$ values.  However, for all models there is a noticeable bias in the approximate posterior for $\sigma_w$.  We also run the unbiased version of Section \ref{subsec:unbiased} with the correlated extension, again for $N=25$.  The same figures in the appendix demonstrate that ueMCMC is able to reduce the bias in the approximate posterior for $\sigma_w$.  The largest difference between the results for $N=1000$ and $N=25$ occurs for the theta-logistic model.  For $N=25$, the unbiased version seems to offer some correction for $\theta_w$ and $\theta_e$ but produces similar results to the biased version for the other parameters.   We find that with the unbiased version the ESS remains similar, but the overall efficiency is slightly reduced.  The reduction in computational efficiency mainly comes here from the extra time to compute the unbiased multivariate normal density estimator (the ESS is roughly the same).  We note that for applications where simulating the transition density consumes the majority of the computation, the additional time associated with computing the unbiased estimator will be significantly less noticeable.

Finally, we investigate improvements that can be obtained in this example when using the RQMC extension.  Here we use $N=50$ ensemble members for each model.   It is evident that the RQMC extension is producing similar marginal posteriors compared to eMCMC with $N=1000$ particles (see Appendix B).  The ESS values for the four models are roughly 2500, 450, 1700, and 2100, which are competitive with standard eMCMC using a significantly larger, $N=200-250$, number of particles (see Table \ref{tab:ecol_efficiency}).  However, the ESS/Time scores for the four models are only roughly 900, 130, 500 and 620 with the RQMC extension.   Given that simulation of the transition density is trivial in this example, the cost associated with generating the RQMC samples is significant and consequently the ESS/time score with the RQMC extension is substantially reduced.  However, in complex examples where simulating the transition density is expensive, the cost associated with RQMC will be far less noticeable.

\subsection{Lorenz Example} \label{subsec:lorenz}

\subsubsection{Model and inference task}

The Lorenz 63 dynamical system \citep{Lorenz1963} is a classic low dimensional example of chaotic behaviour.
An It\^o stochastic differential equation (SDE) version from \cite{Vrettas2015} is
\begin{align*}
dX_t &= \alpha(X_t, \theta) dt + \Sigma^{1/2} dW_t, \\
\alpha(X_t, \theta) &= 
\begin{pmatrix}
\theta_1(X_{2,t} - X_{1,t}) \\
\theta_2 X_{1,t} - X_{2,t} - X_{1,t} X_{3,t} \\
X_{1,t} X_{2,t} - \theta_3 X_{3,t}
\end{pmatrix}, \\
\Sigma &= \begin{pmatrix}
\sigma_1^2 & 0 & 0 \\
0 & \sigma_2^2 & 0 \\
0 & 0 & \sigma_3^2
\end{pmatrix}.
\end{align*}
Here $X_t$ is a vector of the random variables $X_{1,t}, X_{2,t}, X_{3,t}$, and $W_t$ is a vector of three standard uncorrelated Brownian motion processes.
Note that $\Sigma^{1/2}$ is interpreted as a matrix square root.
We assume independent observations $Y_{i,t} \sim \mathcal{N}(X_{i,t}, \sigma_{\text{obs}}^2)$ are made at a grid of prespecified $t$ values for $i=1,2,3$.

Exact simulation of SDEs is extremely challenging, so it is common to work with an Euler-Maruyama discretisation \citep[see e.g.][]{Wilkinson2018}.
For the Lorenz model above this gives,
\[
x_{i+1} = x_i + \alpha(x_i, \theta) \Delta t + \Sigma^{1/2} \sqrt{\Delta t} z_{i+1}, \\
\]
where each $z_{i+1}$ is an independent $\mathcal{N}(0,I_3)$ realisation.
Then $x_i$ is an approximation to $X_t$ for $t=i \Delta t$.

\begin{figure}[tbhp]
	\begin{center}
		\includegraphics[width=0.8\textwidth]{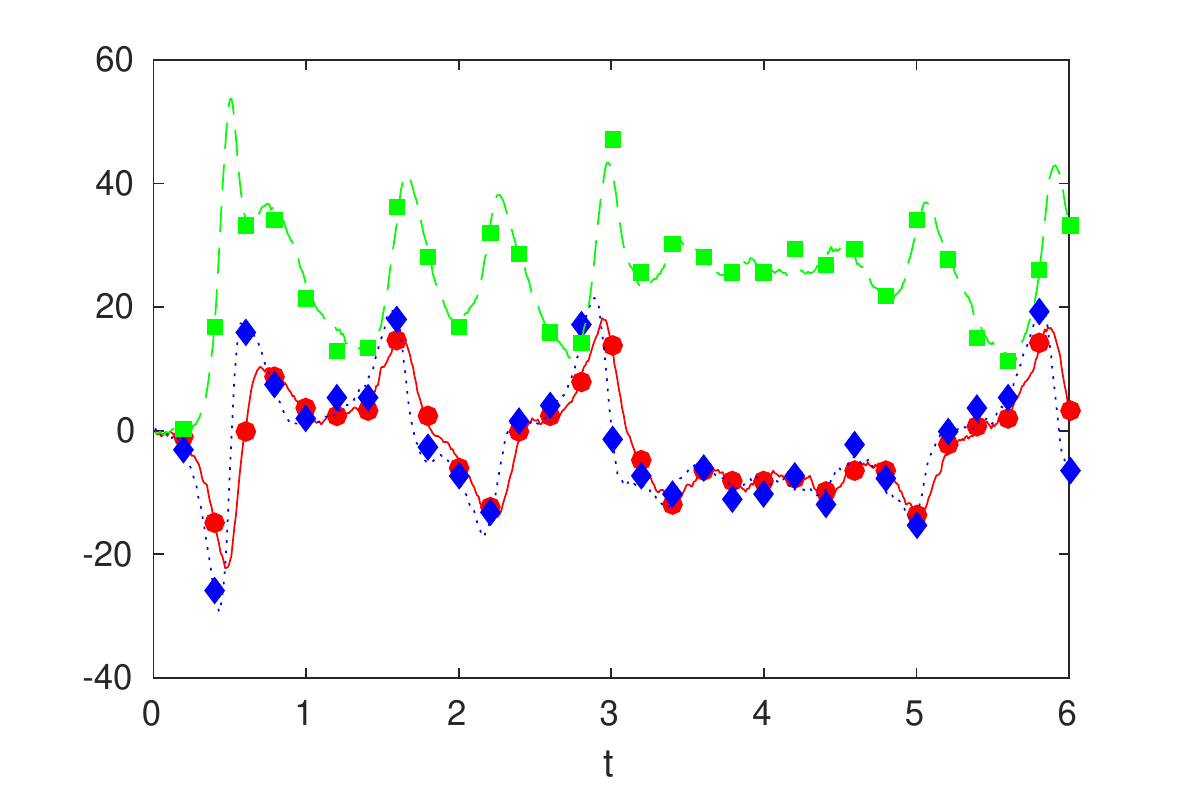}
	\end{center}
	\caption{Simulated Lorenz 63 data.
		The lines show simulated $x_i$ values from the discretised SDE, and the points noisy $y_i$ observations.
		Component $x_{1,i}$ is represented by red circles, $x_{2,i}$ by blue diamonds and $x_{3,i}$ by green squares.}
	\label{fig:DataLorenz}
\end{figure}

Following \cite{Vrettas2015} we simulate data from this discretised model under $\theta = (10,28,8/3)$, $\sigma_i^2=10$ for $i=1,2,3$, $\sigma_{\text{obs}}^2=2$ and $\Delta t = 0.01$.
The initial conditions are $x_0 = (0,0,0)$.
We make observations at $i=20,40,\ldots,600$, corresponding to $t=0.2,0.4,\ldots,6$.
Figure \ref{fig:DataLorenz} shows our data.

\subsubsection{Log likelihoods}

First we compare log-likelihood estimates produced by the EnKF and BPF.
We run each method 5 times for $\theta_1 = 1, 2, \ldots, 20$.
The other parameters are held constant at their true values.
We use 100 particles for both filtering methods.
The average run-times were roughly half as long for EnKF -- $0.019$s -- compared to BPF $0.046$s.

\begin{figure}[tbhp]
	\begin{center}
		\includegraphics[width=0.49\textwidth]{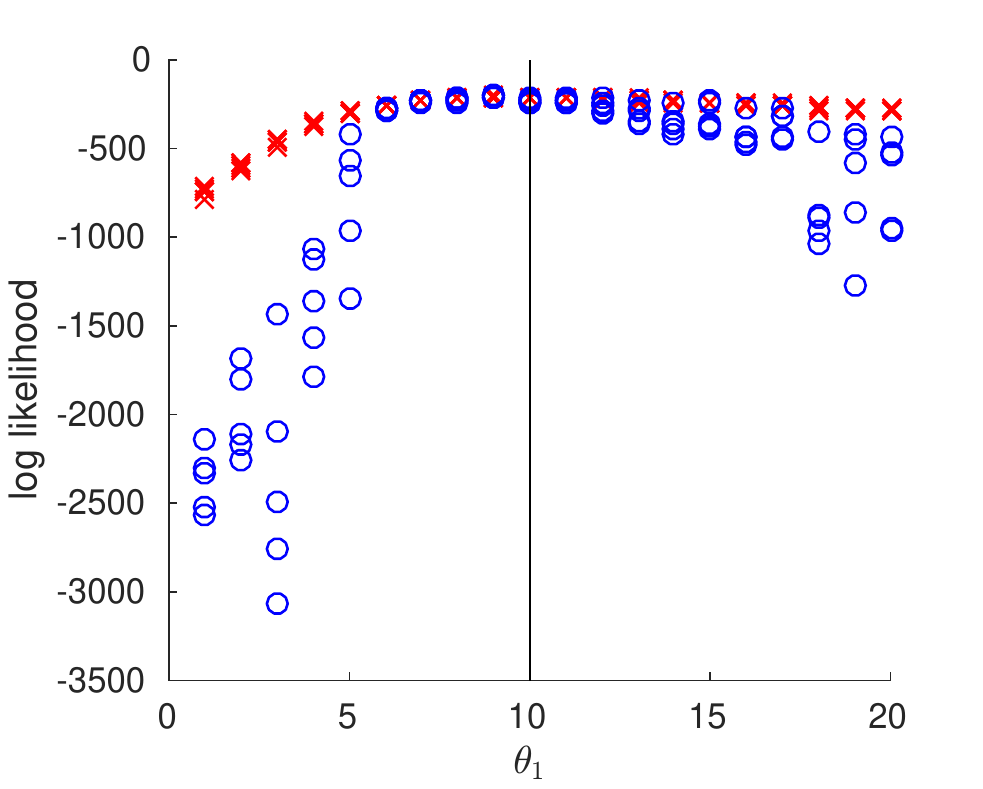}
		\includegraphics[width=0.49\textwidth]{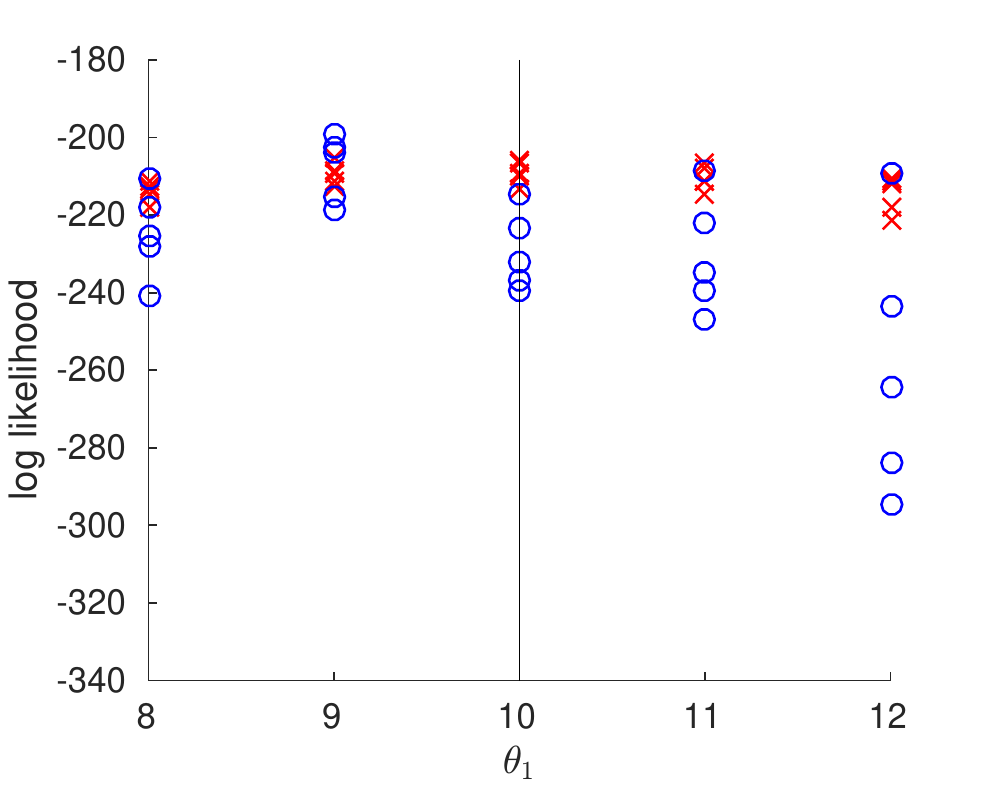}
	\end{center}
	\caption{Lorenz log likelihood estimates using the BPF (blue circles) and EnKF (red crosses) as $\theta_1$ is varied and the other parameters are held constant at their true values.
		Both plots show the same estimates, but the right hand plot zooms in to a smaller plot range.}
	\label{fig:LorenzLogLike}
\end{figure}

Figure \ref{fig:LorenzLogLike} shows the results.
For any $\theta_1$ value, the log likelihood estimates are more variable under BPF than EnKF.
Variability becomes particularly large under BPF when $\theta_1$ is far from its true value.
Such high variance is problematic in pMCMC, as it is likely to cause chains to become stuck.

Figure \ref{fig:LorenzLogLike} suggests that the EnKF and BPF produce similar expected likelihood estimates when $\theta_1$ is close to its true value.
It is hard to draw any conclusions for other $\theta_1$ values, as the BPF expected likelihood will be strongly driven by the upper tail of its log-likelihood estimates, and this would take a very large number of simulations to estimate well.
%WE COULD CONCEIVABLY INVESTIGATE BIASEDNESS WITH A SIMULATION STUDY FOR A $\theta_1$ VALUE CLOSE TO 10

\subsubsection{Inference}

Here we assume $\sigma_{\text{obs}}$ is known, and attempt to infer $\theta_i$ and $\sigma_i$ for $i=1,2,3$.
We assume these parameters have independent exponential prior distributions with rate 0.1.  We ran the EnKF and BPF at the true parameter for 30 times for various choices of $N$ and calculated empirical variances.
Based on these values we select $N=500$ for eMCMC and $N=2500$ for pMCMC.

%We ran three inference methods:
%pMCMC using EnKF likelihood estimates,
%eMCMC using BPF likelihood estimates,
%and the particle ensemble Kalman filter DESCRIPTION TO BE ADDED!  We tuned our MCMC algorithms as discussed in Section \ref{sec:tuning}.
%In particular, we selected the number of particles $N$ for the EnKF and BPF to achieve log-likelihood variances of roughly 1.5 at a representative parameter values:
%rather than marginal posterior medians we used the true parameter values since they are available in this example).

%The pEnKF used 10,000 $\theta$ particles and 100 $x$ particles per ensemble.
%ADD SOME DISCUSSION? ESSENTIALLY THIS IS AN AD-HOC CHOICE.

We ran our algorithms targeting the log transformed parameters.
Both pMCMC and eMCMC achieve acceptance rates in the range 10\% to 20\% indicating reasonable mixing.
Trace plots also suggested good mixing,
with no evidence of chains becoming stuck in the same state for a large number of iterations.
The ESS values for the MCMC outputs were 390 (eMCMC) and 197 (pMCMC).  Run times were 689s and 10,992s for eMCMC and pMCMC respectively.
Interestingly, the pMCMC run time is roughly 15 times that of eMCMC despite using only 5 times as many particles.
%The pEnKF performed well, achieving importance sampling expected sample sizes above 1,000 in most iterations, with a minimum of 119 on its second iteration.  followed by PEnKF (2,500s)

\begin{figure}[tbhp]
	\begin{center}
		\includegraphics[width=\textwidth]{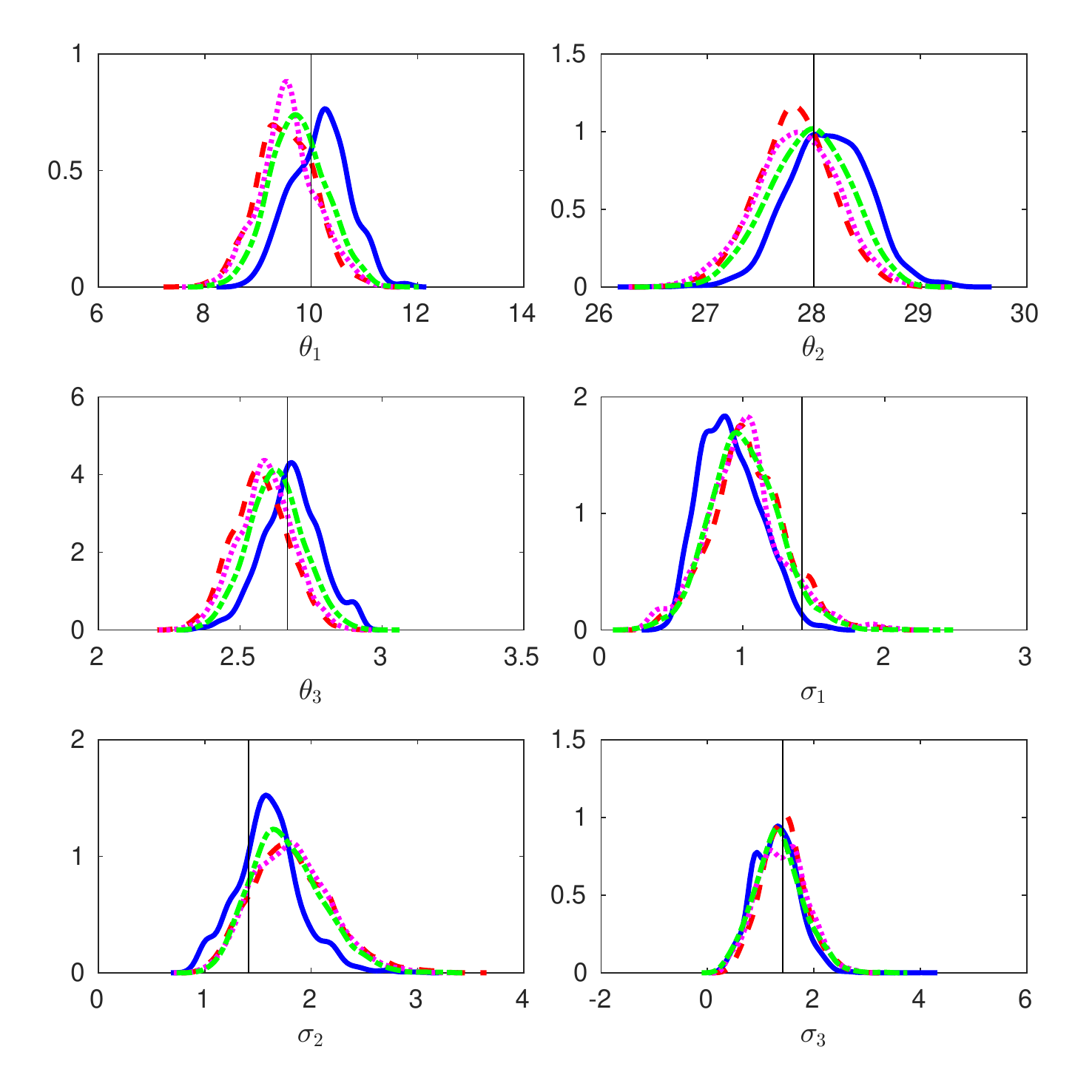}
	\end{center}
	\caption{Estimated Lorenz marginal parameter posteriors using pMCMC (blue solid), eMCMC (red dashed), RQMC eMCMC (magenta dotted), correlated eMCMC (green dot-dash).
		The lines are kernel density estimates based on Monte Carlo samples.
		True parameter values are shown as black vertical lines.}
	\label{fig:LorenzMarginals}
\end{figure}

Figure \ref{fig:LorenzMarginals} shows the resulting marginal posterior estimates.
The eMCMC posterior approximation is similar to the gold standard pMCMC results,
but there are some noticable differences for some parameters
e.g.~the $\theta_1$ and $\theta_3$ posterior marginals are shifted downwards.
Posterior correlations were small for both MCMC methods (all below $0.35$ in magnitude).
%The pEnKF results are a poorer match to the gold-standard posterior, but are reasonably close compared to the prior distribution.
%Also the pEnKF produced some larger posterior correlations, with magnitudes up to $0.69$.

We also ran RQMC and correlated variants of eMCMC
(using $\sigma_u = 0.1$ for the latter).
For RQMC, initial tuning based on variance of the log likelihood selected $N=500$, as for eMCMC.
For correlated eMCMC we used a reduced number of particles, $N=100$.
Posterior marginals are shown in Figure \ref{fig:LorenzMarginals}
and are extremely similar to eMCMC results.
RQMC eMCMC produced a similar acceptance rate and ESS value (305) to eMCMC,
but the cost of QMC sampling increased the run time to 3,073s (roughly a 5 times increase).
Correlated eMCMC increased the acceptance rate (to 26\%) and ESS (to 417) while also reducing the run time (to 195s).

\subsection{Lotka Volterra Example} \label{subsec:lotka}

\subsubsection{Model and inference task} The Lotka-Volterra predator-prey model (e.g.~\citealp{BWK08}) 
describes the continuous time evolution of the non-negative integer-values process 
$X_t=(X_{1,t},X_{2,t})'$ where $X_{1,t}$ denotes prey and $X_{2,t}$ denotes predator. 
Starting from an initial value, $X_t$ evolves according to a Markov jump process 
(MJP) parameterised by stochastic rate constants $c=(c_1,c_2,c_3)'$ and characterised 
by the instantaneous rate or hazard function 
$h(x_t,c)=(h_{1}(x_t,c_1),h_{2}(x_t,c_2),h_{3}(x_t,c_3))'$. Transitions over $(t,t+dt]$ take 
the form of one of three types (prey reproduction, prey death / predator reproduction, 
predator death) with associated probabilities given by
\begin{eqnarray*}
	\textrm{Pr}\left\{X_{1,t+dt}=x_{1,t}+1,X_{2,t+dt}=x_{2,t}|x_{t}\right\} &=& h_{1}(x_t,c_1)dt+o(dt),\\
	\textrm{Pr}\left\{X_{1,t+dt}=x_{1,t}-1,X_{2,t+dt}=x_{2,t}+1|x_{t}\right\} &=& h_{2}(x_t,c_2) dt+o(dt),\\
	\textrm{Pr}\left\{X_{1,t+dt}=x_{1,t},X_{2,t+dt}=x_{2,t}-1|x_{t}\right\} &=& h_{3}(x_t,c_3)dt+o(dt).
\end{eqnarray*}
The hazard function for this system is 
\[
h(X_t,c)=(c_{1}x_{1,t},c_{2}x_{1,t}x_{2,t}, c_{3}x_{2,t})'.
\] 
It is then relatively simple to generate realisations of this process via Gillespie's direct 
method \citep{Gillespie77}, where at time $t$, the dwell time between transition events is drawn from 
an exponential distribution with rate $h_{0}(x_t,c)=\sum_{i=1}^{3}h_i(x_t,c_i)$ and the transition is type 
$i$ with probability proportional to $h_i(x_t,c_i)$. 

We assume that the MJP is observed with Gaussian error so that
\begin{equation*}
Y_t|X_t \sim \mathcal{N}\left\{\left(\begin{array}{c}x_{1,t}\\x_{2,t}\end{array}\right)\,,\,
\left(\begin{array}{cc}\sigma_{1}^{2}&0\\0&\sigma_{2}^{2}\end{array}\right)\right\}.
\end{equation*}
As all parameters of interest must be strictly positive, we consider inference for 
\begin{equation*}
\theta=\left(\log c_{1},\log c_{2},\log c_{3},\log\sigma_1,\log\sigma_2\right)'.
\end{equation*}
We consider two synthetic 
data sets ($\mathcal{D}_1$ and $\mathcal{D}_2$) simulated with rate parameters $c=(0.5,0.0025,0.3)'$ 
and initial condition $x_0=(71,79)'$. We further assume $\sigma_1=\sigma_2=1$ to be unknown. 
To allow the analysis of two data-poor scenarios, data set $\mathcal{D}_1$ has 51 equally spaced observations on $[0,50]$ and data set 
$\mathcal{D}_2$ is constructed by thinning $\mathcal{D}_1$ to give 26 equally spaced observations on $[0,25]$. 

\subsubsection{Inference} We compare the performance of EnKF to the gold standard auxiliary particle filter (APF) 
driven pMCMC scheme described in \cite{GoliWilk15}. In brief, state 
particles are propagated using Gillespie's direct method, with the hazard function replaced by an approximate 
conditioned hazard, derived from a linear Gaussian approximation to the MJP. Full details of this approach, including 
the calculation of the particle filter weights can be found in \cite{GoliWilk15}. 

We follow the practical advice given 
in Section~\ref{sec:tuning} to choose the number of particles / ensemble members $N$ and the scaling of the innovation variance 
in the random walk proposal distribution. We assume independent uniform $U(-8,8)$ priors for the components of 
$\theta$ and ran both eMCMC and pMCMC for $10^5$ iterations. Since the EnKF treats the state as continuous, eMCMC 
used a reflecting barrier at 0 to avoid the state of the system 
going negative.  

\begin{figure}[t]
	\centering
	\includegraphics[angle=0,width=\textwidth]{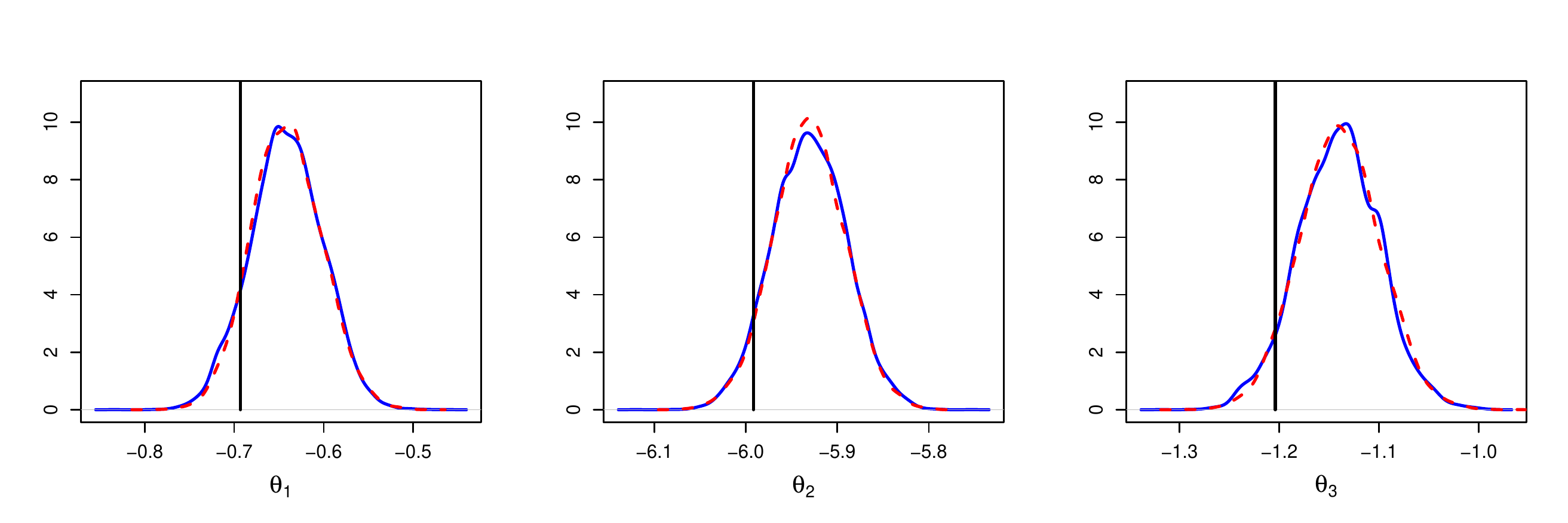}
	\includegraphics[angle=0,width=\textwidth]{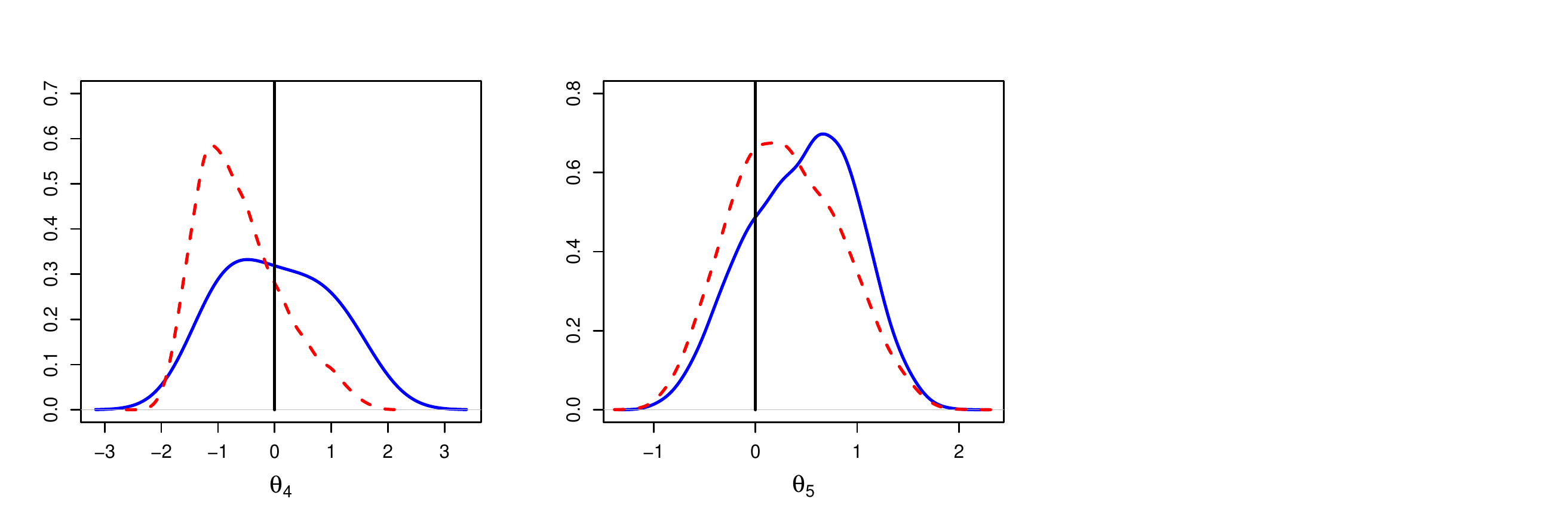}
	\caption{Lotka Volterra data set $\mathcal{D}_1$. Marginal posterior densities based on the output of pMCMC (solid) and eMCMC (dashed).}\label{fig:figLV}
\end{figure}

\begin{figure}[t]
	\centering
	\includegraphics[angle=0,width=\textwidth]{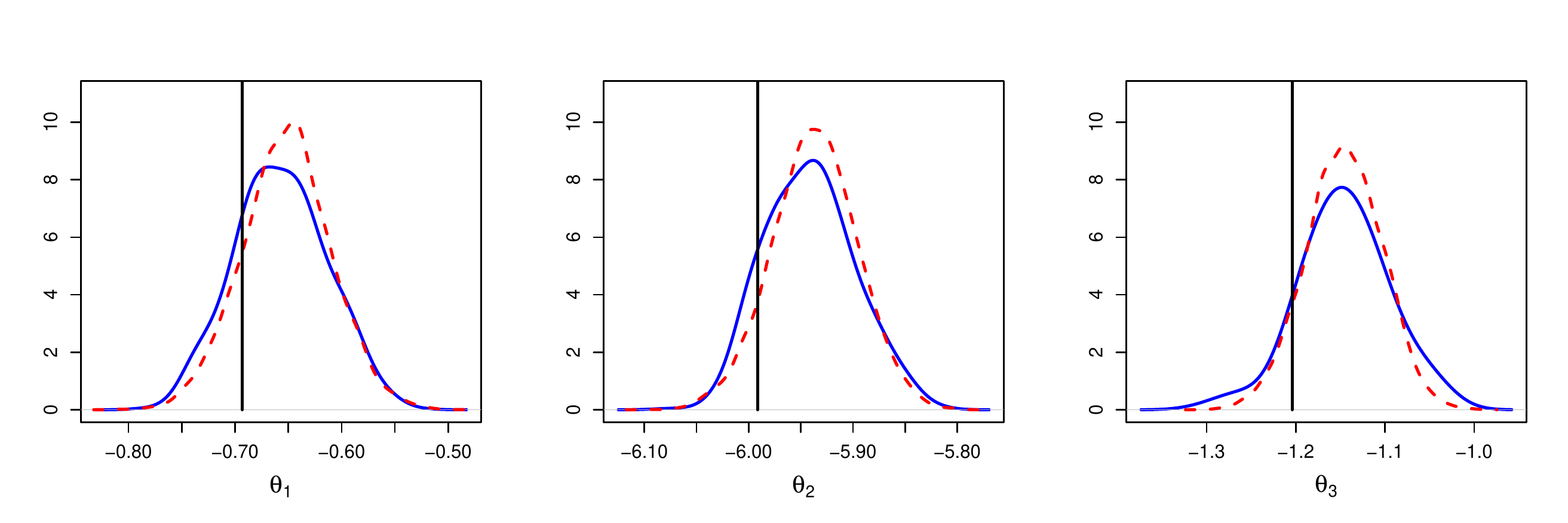}
	\includegraphics[angle=0,width=\textwidth]{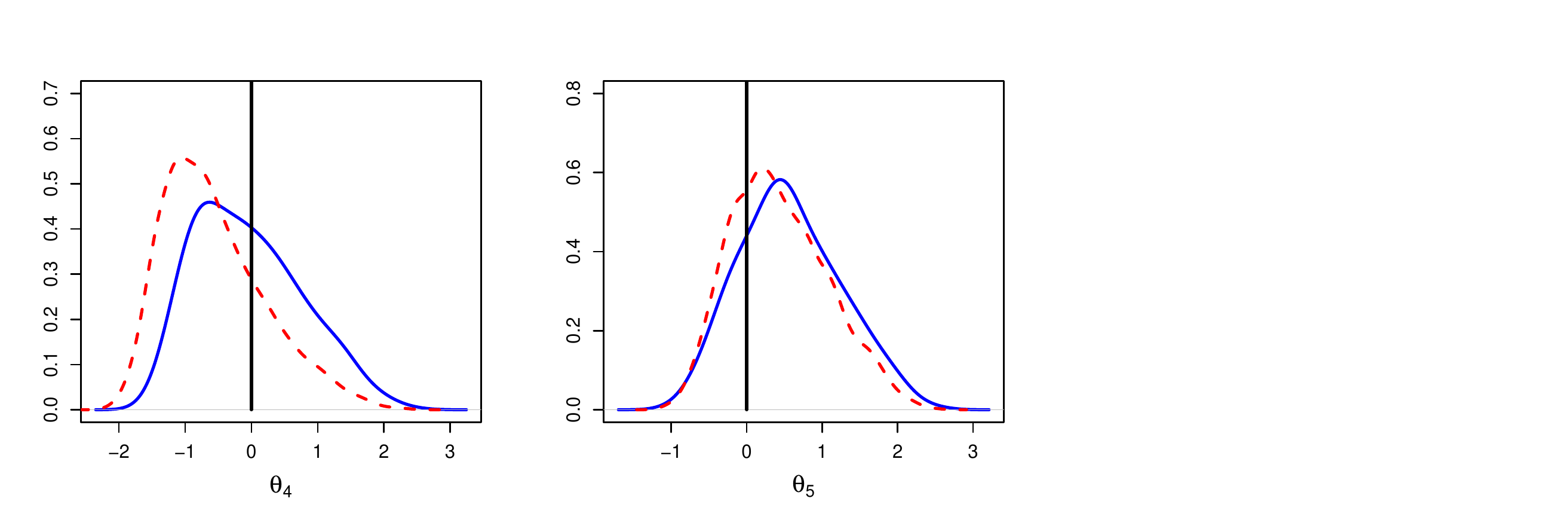}
	\caption{Lotka Volterra data set $\mathcal{D}_2$. Marginal posterior densities based on the output of pMCMC (solid) and eMCMC (dashed).}\label{fig:figLV2}
\end{figure}

\begin{figure}[t]
	\centering
	\includegraphics[angle=0,width=\textwidth]{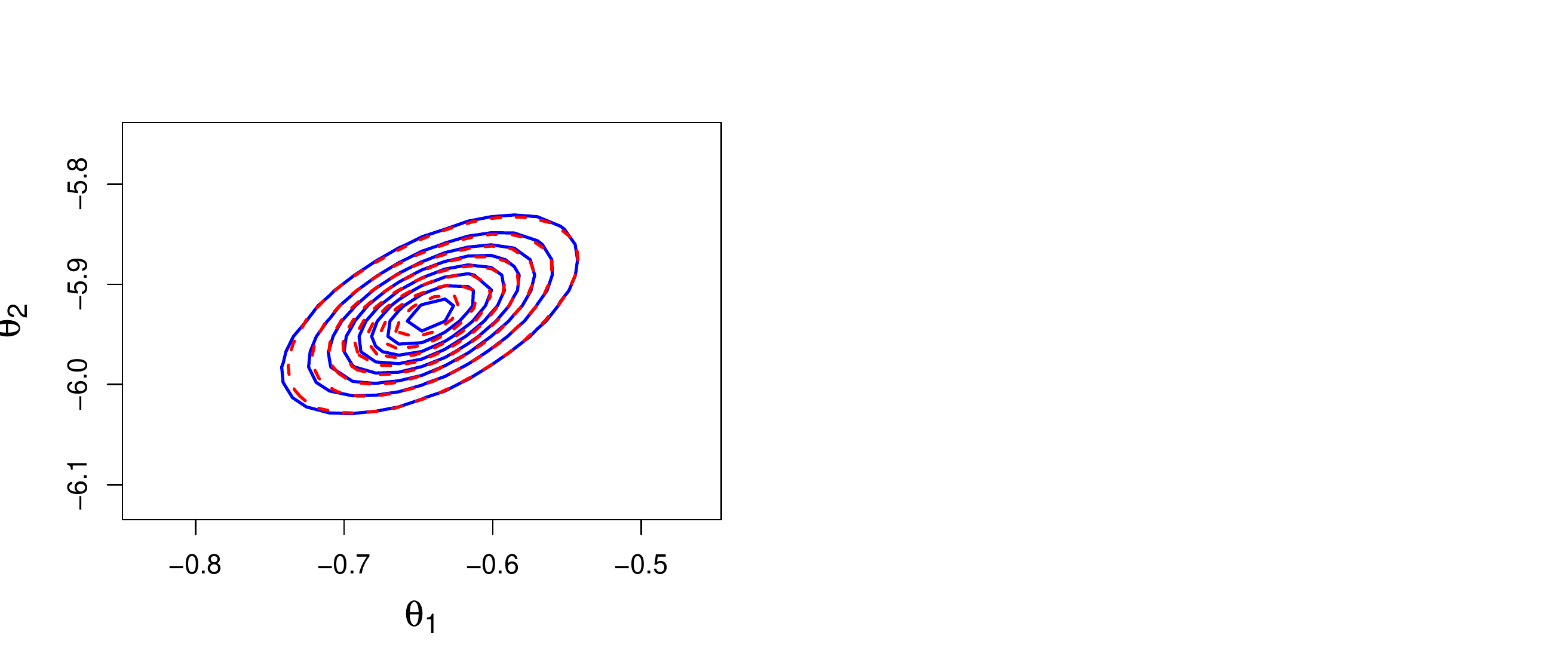}
	\includegraphics[angle=0,width=\textwidth]{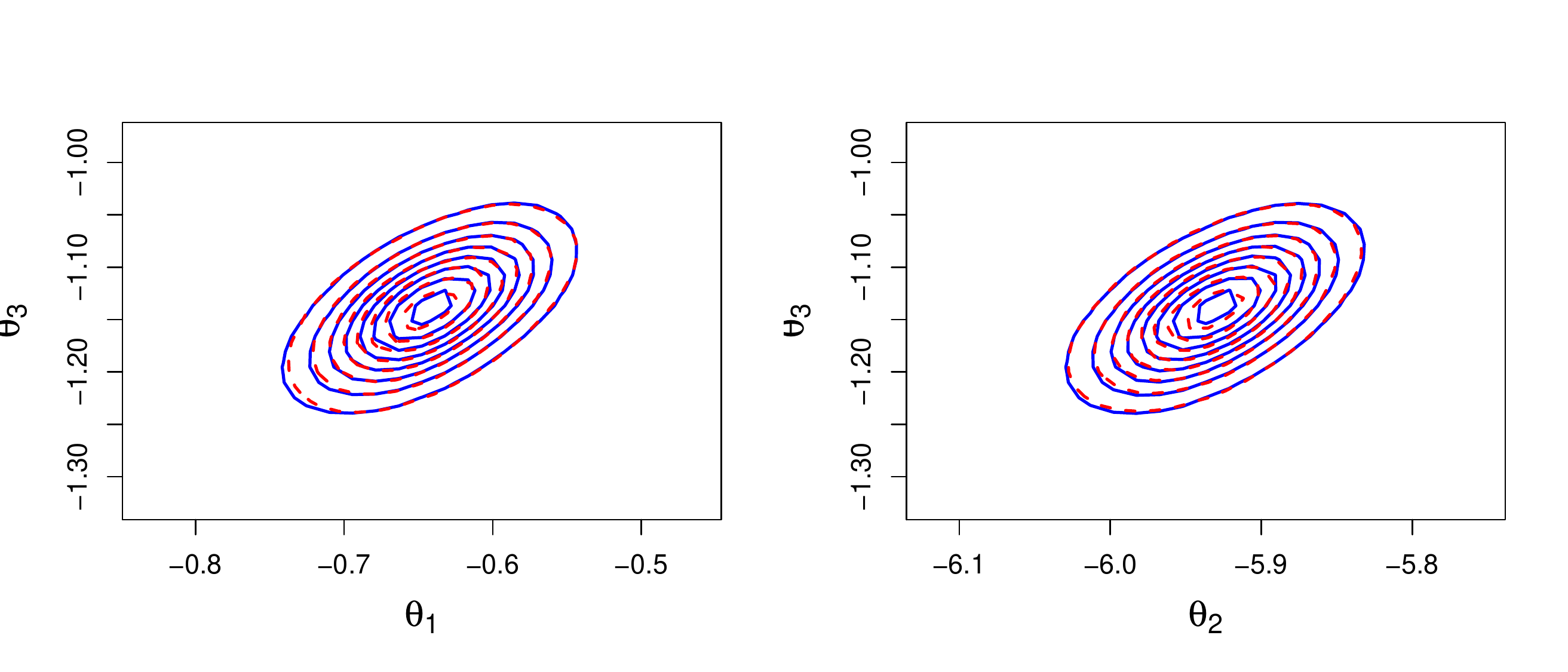}
	\caption{Lotka Volterra data set $\mathcal{D}_1$. Bivariate posterior densities based on the output of pMCMC (solid) and eMCMC (dashed).}\label{fig:figLV3}
\end{figure}

The results are summarised by Table~\ref{tab:tabLV} and Figures~\ref{fig:figLV}--\ref{fig:figLV3}. 
We see that for both data sets, the output of eMCMC is consistent with the true values that produced the 
data and, more importantly, the ground truth posterior based on the output of pMCMC. For data set 
$\mathcal{D}_1$, eMCMC required more particles than pMCMC but gives better overall efficiency 
(as measured by the ESS per second) since sampling from the propagation construct in 
the auxiliary particle filter is relatively expensive. We see an increase of about a factor of 3. For data set $\mathcal{D}_2$, the number 
of particles required by pMCMC must be increased, since the propagation construct is based on a linear Gaussian approximation 
of the true (but unknown) hazard function of the conditioned MJP. The construct breaks down as observations are made 
sparsely in time (and the dynamics of the conditioned process are nonlinear between observations). Ensemble MCMC on the 
other hand seems to work well, requiring even fewer particles than for $\mathcal{D}_1$. We see an increase 
in overall efficiency (compared to pMCMC) of a factor of around 55. 

\begin{table}[t]
	\begin{center}
		\begin{tabular}{|l|cccccc|}
			\hline
			Filter & $N$ & $\tau$ & Acc. rate& ESS & Time (s)  & ESS$/$Time\\
			\hline	
			& \multicolumn{6}{|c|}{$\mathcal{D}_{1}$ (51 obs. every 1 time unit)} \\
			APF   &55  &1.4 &0.11 &1117 &26298 &0.042  \\
			EnKF  &150 &1.4 &0.11 &1762 &12299 &0.143  \\  
			\hline
			& \multicolumn{6}{|c|}{$\mathcal{D}_{2}$ (26 obs. every 2 time units)} \\
			APF   &350  &1.4 &0.08 &1156 &165015 &0.007  \\
			EnKF  &65   &1.4 &0.11 &2054 &5282   &0.389  \\ 
			\hline
		\end{tabular}
		\caption{Summaries for the Lotka Volterra application: number of particles $N$, standard deviation of the noise in the log-posterior ($\tau$) at the posterior median, 
			acceptance rate, multivariate effective sample size (ESS), wall clock time in seconds 
			and ESS per second.}\label{tab:tabLV}
	\end{center}
\end{table}

\subsection{Autoregulatory Network Example} \label{subsec:Autoregulatory}

\subsubsection{Model and inference task} A commonly used mechanism for
auto-regulation in prokaryotes which has been well-studied and
modelled is a negative feedback mechanism whereby dimers of a protein
repress its own transcription \citep[e.g.][]{ARM98}. A 
simplified model for such a prokaryotic auto-regulation, 
based on this mechanism of dimers of a protein coded 
for by a gene repressing its own transcription into RNA, can be 
found in \cite{GoliWilk05} \cite[see also][]{GoliWilk11}. 

Let $X_t=(X_{1,t},X_{2,t},X_{3,t},X_{4,t},X_{5,t})'$ 
denote the number of copies of the unbound gene $X_{1,t}$, 
bound gene $X_{2,t}$, RNA $X_{3,t}$, protein $X_{4,t}$ and dimers of the protein 
$X_{5,t}$. We assume that $X_t$ evolves according to a Markov 
jump process. The possible transitions can be succinctly described by the 
pseudo-reaction list
\begin{align*}
\mathcal{R}_1: & \quad X_1+X_5 \longrightarrow X_2   & \mathcal{R}_2: & \quad X_2 \longrightarrow X_1+X_5   \\
\mathcal{R}_3: & \quad X_1 \longrightarrow X_1 + X_3 & \mathcal{R}_4: & \quad X_3 \longrightarrow X_3 + X_4 \\
\mathcal{R}_5: & \quad 2X_4 \longrightarrow X_5      & \mathcal{R}_6: & \quad X_5 \longrightarrow 2X_4      \\
\mathcal{R}_7: & \quad X_3 \longrightarrow \emptyset & \mathcal{R}_8: & \quad X_4 \longrightarrow \emptyset
\end{align*}
where, for example, occurence of $\mathcal{R}_1$ at time $t$ 
reduces $X_{1,t}$ and $X_{5,t}$ by 1, increases $X_{2,t}$ 
by 1, and leaves the remaining components unchanged. The associated 
hazard function is
\[
h(x_t,c)=(c_1 x_{1,t}x_{5,t}, c_2 x_{2,t},
c_3 x_{1,t}, c_4 x_{3,t}, c_5 x_{4,t}(x_{4,t}-1)/2,
c_6 x_{5,t}, c_7 x_{3,t}, c_8 x_{4,t})'.
\]

We consider here two challenging synthetic data sets, each consisting of 101 observations at integer times on $X_{3,t}$ (RNA) and total protein 
counts, $X_{4,5}+2X_{5,t}$ so that $X_{1,t}$, $X_{2,t}$, $X_{4,t}$ and $X_{5,t}$ are not observed exactly. Moreover, as in 
Section~\ref{subsec:lotka}, we corrupt the observations by adding independent, Gaussian $\mathcal{N}\{0,\textrm{diag}(\sigma_{1}^2,\sigma_{2}^2)\}$ 
innovations to each count. We fix $\sigma_1=\sigma_2=1$ for data set $\mathcal{D}_1$ and $\sigma_1=\sigma_2=0$ for data set $\mathcal{D}_2$. 
Following \cite{GoliWilk05}, we use the rate constants
\[
c=(0.1,0.7,0.35,0.2,0.1,0.9,0.3,0.1)'.
\]
We assume that the initial condition $x_0=(5,5,8,8,8)'$, the measurement 
error variances and the rate constants of the reversible dimerisation 
reactions ($c_5$ and $c_6$) are known leaving $\theta_{i}=\log c_i$, 
$i=1,2,3,4,7,8$ as the object of inference.

\subsubsection{Inference} We again compare the performance of eMCMC to the gold standard auxiliary particle filter 
driven pMCMC scheme described in \cite{GoliWilk15}. The number of particles / ensemble members $N$ 
was chosen as in Section~\ref{sec:tuning}. We assign independent Gamma $Ga(1,0.5)$ priors to 
each unknown rate constant and ran eMCMC and pMCMC for $2\times 10^5$ iterations. Note that when running eMCMC 
for data set $\mathcal{D}_2$, the values of $\sigma_1$ and $\sigma_2$ result in no shifting of the ensemble members, 
rendering this step ineffectual. We therefore modified eMCMC for this scenario by setting the measurement error 
variance to be ``small'' throughout the algorithm's execution. Specifically, we found that setting $\sigma_1^2=\sigma_2^2=0.01$ 
gave reasonable mixing, at the expensive of introducing additional bias into the eMCMC posterior. 

\begin{figure}[t]
	\centering
	\includegraphics[angle=0,width=\textwidth]{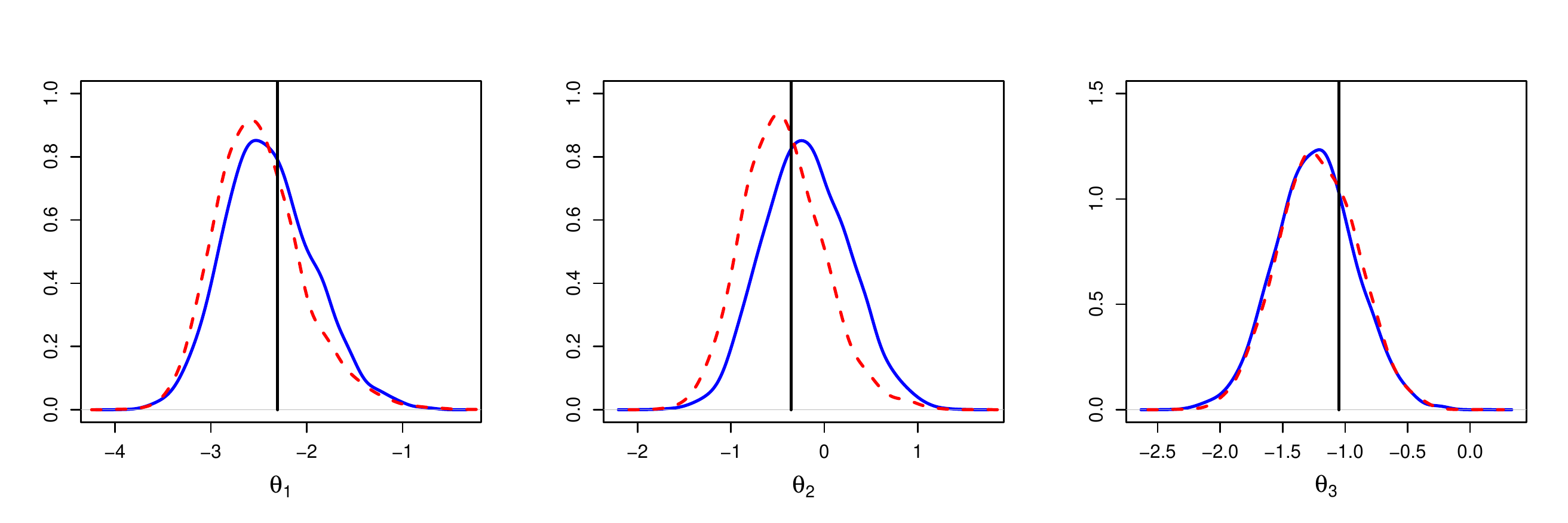}
	\includegraphics[angle=0,width=\textwidth]{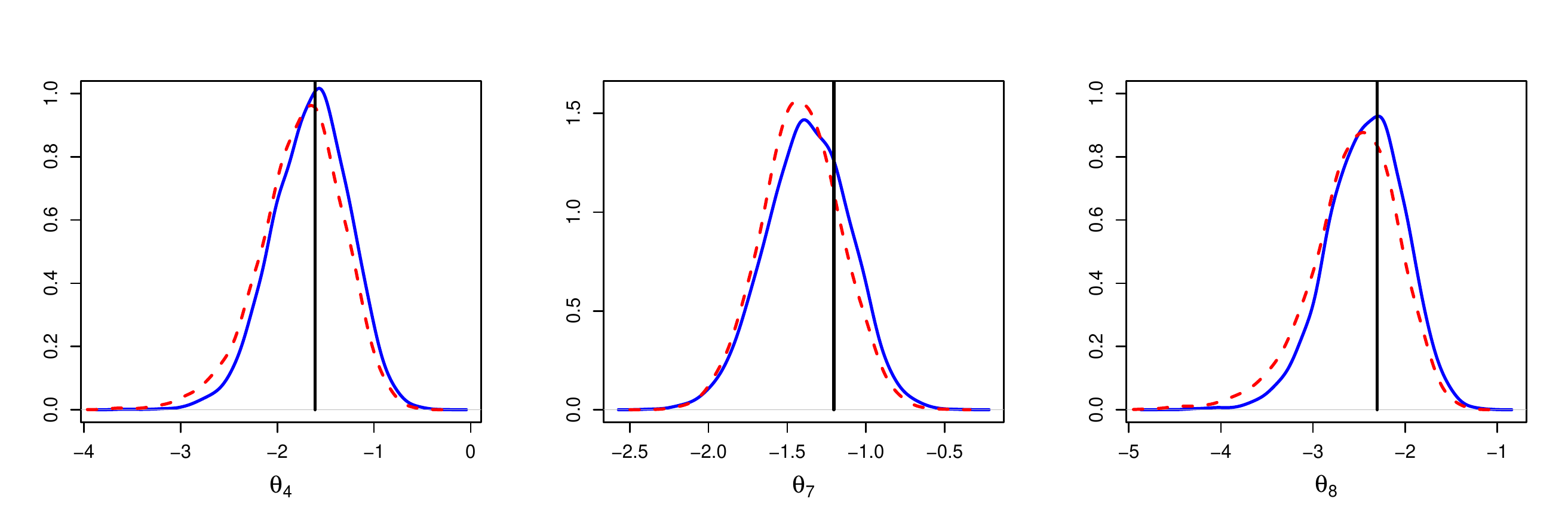}
	\caption{Autoregulatory data set $\mathcal{D}_1$. Marginal posterior densities based on the output of pMCMC (blue solid) and eMCMC (red dash).}\label{fig:figAR}
\end{figure}

\begin{figure}[t]
	\centering
	\includegraphics[angle=0,width=\textwidth]{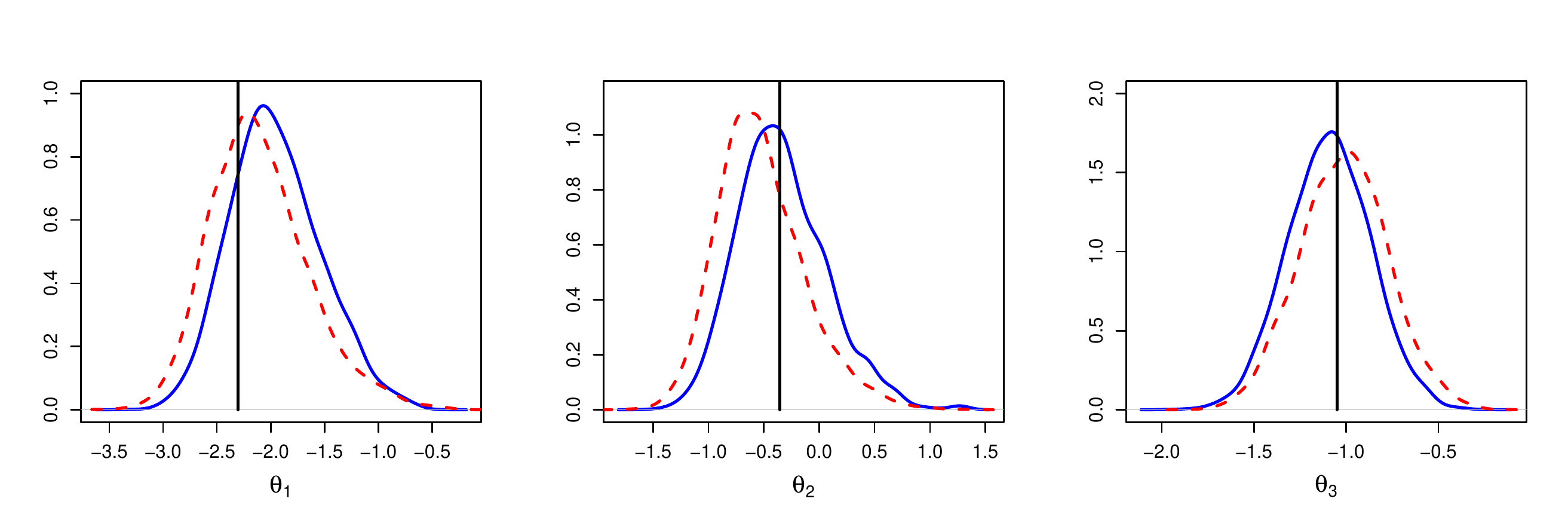}
	\includegraphics[angle=0,width=\textwidth]{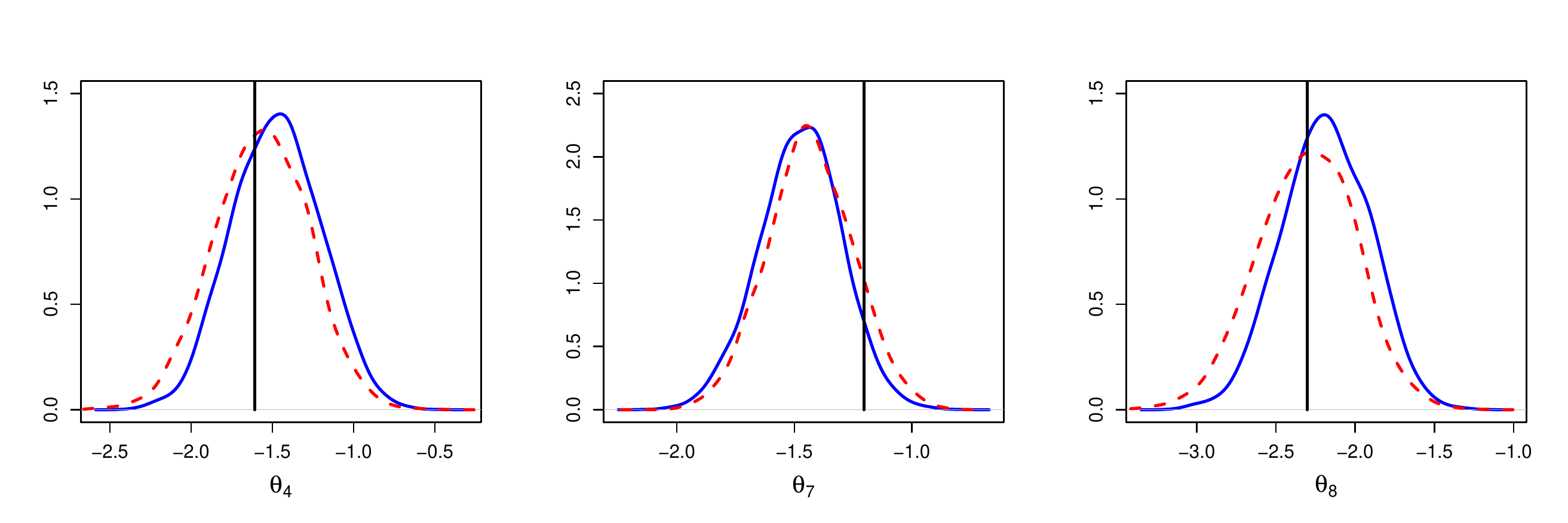}
	\caption{Autoregulatory data set $\mathcal{D}_2$. Marginal posterior densities based on the output of pMCMC (blue solid) and eMCMC (red dash).}\label{fig:figAR2}
\end{figure}

Table~\ref{tab:tabAR} and Figures~\ref{fig:figAR}--\ref{fig:figAR2} summarise the results. It is clear that eMCMC gives 
output that is consistent with the true values that produced the data and output from pMCMC, which exactly targets the 
posterior of interest. We therefore compare overall efficiency of eMCMC and pMCMC in terms of effective sample size per second, 
as reported in Table~\ref{tab:tabAR}. For data set $\mathcal{D}_1$, eMCMC requires half the number of particles of pMCMC and gives 
a comparable ESS value. In terms of overall efficiency, eMCMC outperforms pMCMC by around a factor of 4. For data set $\mathcal{D}_2$, 
pMCMC requires around 2000 particles, due to the strict requirement of particle trajectories having to ``hit'' the observations 
to recieve a non zero weight. Ensemble MCMC on the other hand is able to give a comparable ESS value with just 370 particles. Consequently, for this example, eMCMC outperforms pMCMC by around a factor of 11.   

\begin{table}[t]
	\begin{center}
		\begin{tabular}{|l|cccccc|}
			\hline
			Filter   & $N$ & $\tau$ & Acc. rate& ESS & Time (s)  & ESS$/$Time\\
			\hline	
			& \multicolumn{6}{|c|}{$\mathcal{D}_{1}$ ($\sigma_1=\sigma_2=1$)} \\
			APF   &400  &1.3 &0.15 &3348 &72081 &0.046  \\
			EnKF  &200  &1.5 &0.13 &2972 &15862 &0.187  \\  
			\hline
			& \multicolumn{6}{|c|}{$\mathcal{D}_{2}$ ($\sigma_1=\sigma_2=0$)} \\
			APF   &2000  &1.4 &0.11 &3314 &403456 &0.0082  \\
			EnKF  &370   &1.4 &0.11 &3176 &34112  &0.0931  \\  
			\hline
		\end{tabular}
		\caption{Summaries for the autoregulatory example: number of particles $N$, standard deviation of the noise in the log-posterior ($\tau$) at the posterior median, 
			acceptance rate, multivariate effective sample size (ESS), wall clock time in seconds 
			and ESS per second.}\label{tab:tabAR}
	\end{center}
\end{table}

\subsection{Neuroscience Example} \label{subsec:neuroscience}

\subsubsection{Model}

We investigate the following realistic Neural Population Model (NPM) for brain
activity. This model (see e.g.~\citealp{Bojak2005}) is known as the Liley model,
and Bayesian inference for the parameters of this model has previously
been described by \citet{Maybank2017}. Here a high-level description of the model
is presented; more detail can be found in this latter paper. The model consists
of the following differential equations, where $k=e,i$ for excitatory
and inhibitory contributions: 
\begin{align}
\left(\frac{d}{d t}+\gamma_{ek}\right)\left(\frac{d}{d t}+\bar{\gamma}_{ek}\right)I_{ek}(t) & =\exp\left(\gamma_{ek}d_{ek}\right)\Gamma_{ek}\tilde{\gamma}_{ek}\left[N_{ek}^{\beta}S_{e}\left(h_{e}(t)\right)+\Phi_{ek}(t)+\bar{p}_{ek}+\delta_{ek}p(t)\right],\\
\left(\frac{d}{d t}+\gamma_{ik}\right)\left(\frac{d}{d t}+\bar{\gamma}_{ik}\right)I_{ik}(t) & =\exp\left(\gamma_{ik}d_{ik}\right)\Gamma_{ik}\tilde{\gamma}_{ik}\left[N_{ik}^{\beta}S_{i}\left(h_{i}(t)\right)\right],\\
\left(\frac{d}{d t}+v\Lambda\right)^{2}\Phi_{ek}(t) & =v^{2}\Lambda^{2}N_{ek}^{\alpha}S_{e}\left(h_{e}(t)\right),\\
\Phi_{ek}(t) & =0,
\end{align}
where the Kronecker delta $\delta_{ek}$ admits only excitatory noise
input $p$ (white noise with zero mean and fixed standard deviation) to this stochastic differential equation system, and where $S$ is a sigmoidal activation function.

We model an electroencephalogram (EEG) time-series as noisy observations
of the $\ensuremath{h_{e}}$ variable of this NPM, assuming that the
EEG observations are linearly proportional to $h_{e}$ with some added
observational noise.
\begin{equation}
y_{i}=h_{e}(i\cdot\Delta t)+z_{i}
\end{equation}
where $\Delta t$ is some constant time-step and the $z_{i}$ are
iid normal random variables $z_{i}\sim\mathcal{N}\left(0,\sigma^{2}\right)$
for $i=0,\ldots,n-1$.

In this paper an input noise of variance $10^{8}$ was used to simulate
data, and the dynamics (consisting of 14 state variables) were simulated using the Euler-Maruyama method
with step size $2.5\times 10^{-3}$. Table \ref{t:liley_params} gives the prior distributions for the parameters that were treated as unknown, giving the uniform priors that restrict the parameters to ranges found to be plausible in \citet{Bojak2005}; other parameters were fixed to values chosen from the ranges given by \citet{Bojak2005}.

\begin{table}
	\centering
\begin{tabular}{|c|c|c|}
\hline 
Parameter & Prior & Value for simulation\tabularnewline
\hline 
\hline 
$\ensuremath{\Gamma_{ee}}$ & $\mathcal{U}\left(0.1,2\right)$ & 0.10631\tabularnewline
\hline 
$\ensuremath{\Gamma_{ei}}$ & $\mathcal{U}\left(0.1,2\right)$ & 0.64105\tabularnewline
\hline 
$\ensuremath{\Gamma_{ie}}$ & $\mathcal{U}\left(0.1,2\right)$ & 0.46477\tabularnewline
\hline 
$\ensuremath{\Gamma_{ii}}$ & $\mathcal{U}\left(0.1,2\right)$ & 0.28663\tabularnewline
\hline 
$\gamma_{ee}$ & $\mathcal{U}\left(100,1000\right)$ & 291.5\tabularnewline
\hline 
$\gamma_{ei}$ & $\mathcal{U}\left(100,1000\right)$ & 697.76\tabularnewline
\hline 
$\gamma_{ie}$ & $\mathcal{U}\left(10,500\right)$ & 458.67\tabularnewline
\hline 
$\gamma_{ii}$ & $\mathcal{U}\left(10,500\right)$ & 82.33\tabularnewline
\hline 
$\bar{p}_{ee}$ & $\mathcal{U}\left(0,10000\right)$ & 6603.4\tabularnewline
\hline 
$\bar{p}_{ei}$ & $\mathcal{U}\left(0,10000\right)$ & 2625.7\tabularnewline
\hline 
$\sigma$ & $\mathcal{U}\left(0,10\right)$ & 0.01\tabularnewline
\hline 
\end{tabular}
\caption{Parameters of the Liley model.}
\label{t:liley_params}
\end{table}

\subsubsection{Results}

We compared the performance of eMCMC and pMCMC on data simulated
from the Liley model for parameters that result in quasi-linear dynamics
about a stable fixed point. The work in \citet{Maybank2017} suggests
that the accuracy of the parameter posterior is likely to be improved
by using a method that is suitable for non-linear systems (such as
particle MCMC) compared to using a linearised approach such as the
extended Kalman filter or the approach introduced in \citet{Maybank2017}.
In both MCMC approaches we use a Metropolis-Hastings approach, with
a truncated multivariate normal proposal for the 11 parameters with
covariance chosen to be $2.562^2/11$ times the estimated posterior covariance from pilot
runs (this scaling being recommended by \citealp{Sherlock2015}). We considered a situation that is challenging for a particle
filter, with a relatively small measurement noise of $\sigma=0.01$.

\begin{figure}
	\centering
\includegraphics[scale=0.3]{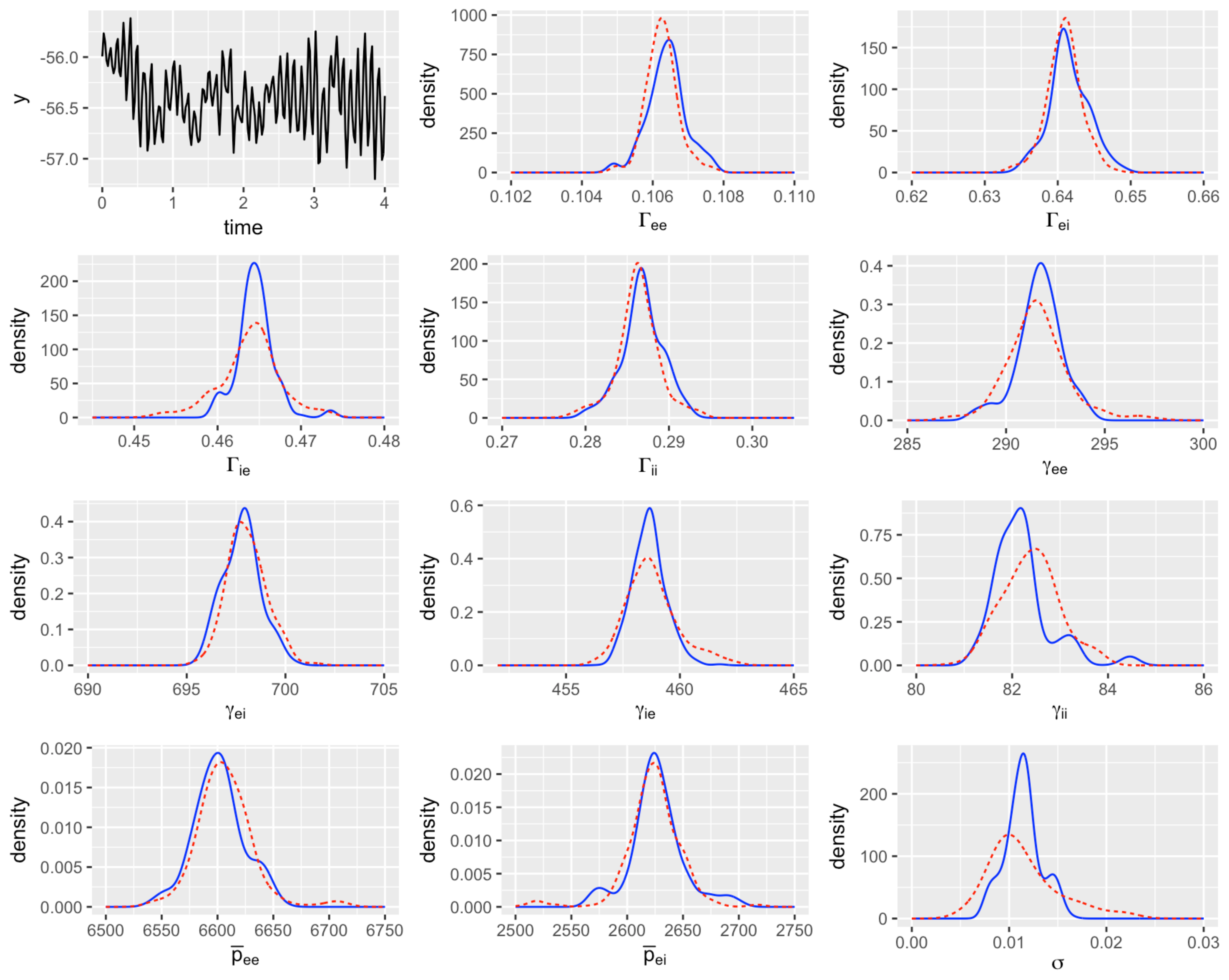}

\caption{Top left: simulated data from the Liley model with added measurement noise. Other plots: estimated marginal posterior distributions for the parameters of the Liley model based on pMCMC (blue solid) and eMCMC (red dash).}
\label{f:neuro_results}
\end{figure}

We study a simulated data set, generated from the model using Euler-Maruyama approximation.
The data, shown in Figure \ref{f:neuro_results} has a length of 4s and a sampling frequency
of 50Hz, and was generated for the parameters $(\ensuremath{\Gamma_{ee}=0.10631},\ensuremath{\Gamma_{ei}}=0.64105,\ensuremath{\Gamma_{ie}}=0.46477,\Gamma_{ii}=0.28663,\ensuremath{\gamma_{ee}=291.5},\ensuremath{\gamma_{ei}}=697.76,\ensuremath{\gamma_{ie}}=458.67,\ensuremath{\gamma_{ii}}=82.33,\bar{p}_{ee}=6603.4,\bar{p}_{ei}=2625.7,\sigma=0.01)$.

We ran 40 chains of $1000$ iterations of pMCMC and eMCMC on this data, all initialised from the parameters at which the data was generated then run for an additional 500 iterations. Based on the scheme described in Section \ref{sec:tuning}, we chose 1000 particles for the BPF in pMCMC, and 100 ensemble members for the EnKF in eMCMC. Both algorithms were implemented with early rejection schemes, as detailed in Section \ref{subsec:rejection}. In both cases the early rejection results in a reduction in computational cost of approximately a factor of two; with this scheme each iteration of pMCMC took an average of 1415s, compared to the average of 91s for eMCMC. The mean acceptance rate for pMCMC was $0.33\%$, compared to $0.93\%$ for eMCMC, indicating that eMCMC is (by this measure) is approximately three times as efficient whilst being more than 15 times faster. Pilot runs on longer simulated time series suggest that the efficiency of eMCMC (relative to pMCMC) improves as the length of the time series increases, but in these cases the computational cost of pMCMC was too large to permit a rigorous comparison. Kernel density estimates of the marginal posterior of each parameter are shown in Figure \ref{f:neuro_results}: we observe that the posteriors obtained by both methods are similar.

\section{Discussion} \label{sec:discussion}

In this paper we replace the BPF with the EnKF within a particle MCMC algorithm. We have demonstrated on a variety of examples that significant computational gains can be achieved without sacrificing much on posterior accuracy. 

If exact posterior inferences are essential, there are likely to be ways to exploit our EnKF approach to improve computational performance. For example, it could be used as the cheap approximate likelihood within a delayed-acceptance MCMC algorithm (e.g.~\citealp{Sherlock2017} and \citealp{Golightly2014}) or importance sampling scheme \citep{franks2017}. Alternatively, we might bridge our approximate posterior with the true posterior using sequential Monte Carlo. Further, our approach could be used in pilot MCMC runs to more quickly identify the regions of the parameter space with non-negligible posterior support and assist MCMC tuning generally.  Particularly in the posterior tails we find that the EnKF likelihood estimator have significantly lower variance than the BPF likelihood estimator.

It is important to note that there will likely be many applications where the EnKF approximation may not be appropriate. The approach relies on being able to approximate the filtering distribution reasonably well with a Gaussian density. However, our paper illustrates that there are a wide class of models where our approach can provide reasonable accuracy.  Further, \citet{Katzfuss2019} present a hierarchical approach for allowing non-Gaussian observation densities with EnKF methods, which would also be applicable to our approach.

We also did some comparisons of our approach with the particle EnKF (pEnKF) method of \citet{Katzfuss2019} (see their Algorithm 4). Briefly,  their approach is a sequential algorithm that evolves a population of static parameters (particles) over time, where each particle has an associated ensemble for the hidden state.  The approach uses the EnKF approximation of the likelihood to re-weight the particles.  The ensemble of latent states is propagated by the transition density and the particles are propagated via resampling and jittering step.  We found this algorithm to be fast.  However, it requires the user to choose several aspects of the algorithm and we found the posterior approximations to be significantly less accurate then what we obtain here.  However, we suggest that our approach could be incorporated into the SMC$^2$ algorithm of \citet{Chopin2013}, which uses an MCMC kernel for jittering particles and thus preserves the current target.  We leave that for further research.

We did not consider posterior inference for the hidden states in this paper.  It might be possible to combine our method with the ensemble Kalman smoother of \citet{VanLeeuwen1996}, but this requires further investigation.

In this paper we compared the most commonly used particle filter (the BPF, except in the Markov jump process examples) and EnKF (the stochastic EnKF) within MCMC algorithms. In future work it would be interesting to compare extensions to both approaches. Extensions to the BPF are familiar to many in computational statistics (e.g. adaptive resampling, MCMC rejuvenation moves, the auxiliary PF) and the improvements they can bring to particle MCMC algorithms are relatively well understood. In the paper we have seen how ideas previously used in the particle MCMC context (i.e. using Quasi Monte Carlo, and the correlated approach) can also be exploited in the EnKF case. Other extensions and alternatives to the stochastic EnKF from the DA literature are also possible and have the potential to provide further improvements in the particle MCMC setting. Examples are: the deterministic EnKF \citep{Tippett2003}, which uses a deterministic rather than a stochastic transformation in the shift step, which may further reduce the variance of the likelihood estimates (but which may introduce further bias for nonlinear models); the equivalent weights particle filter \citep{vanLeeuwen2010}, which uses a deterministic transform in each step of a PF to avoid degeneracy (but which may, again, introduce bias); or for state spaces of high dimension, strategies such as variance inflation and localisation \citep{Katzfuss2016}.

\section*{Acknowledgments}

CD and DP are grateful to RGE for providing support to present at a Bayesian workshop at the University of Reading, where this research project was instigated. RGE thanks Philip Maybank for the implementation of the solver for the Liley model, and Philip Maybank and Ingo Bojak for invaluable discussions about this model.

\bibliographystyle{apalike}
\bibliography{Refs}

\newpage

\section*{Appendix A - Correlated and unbiased results for the population ecology example} \label{app:corr_results}

\begin{figure}[tbhp]
	\begin{center}
		\includegraphics[width=0.7\textwidth]{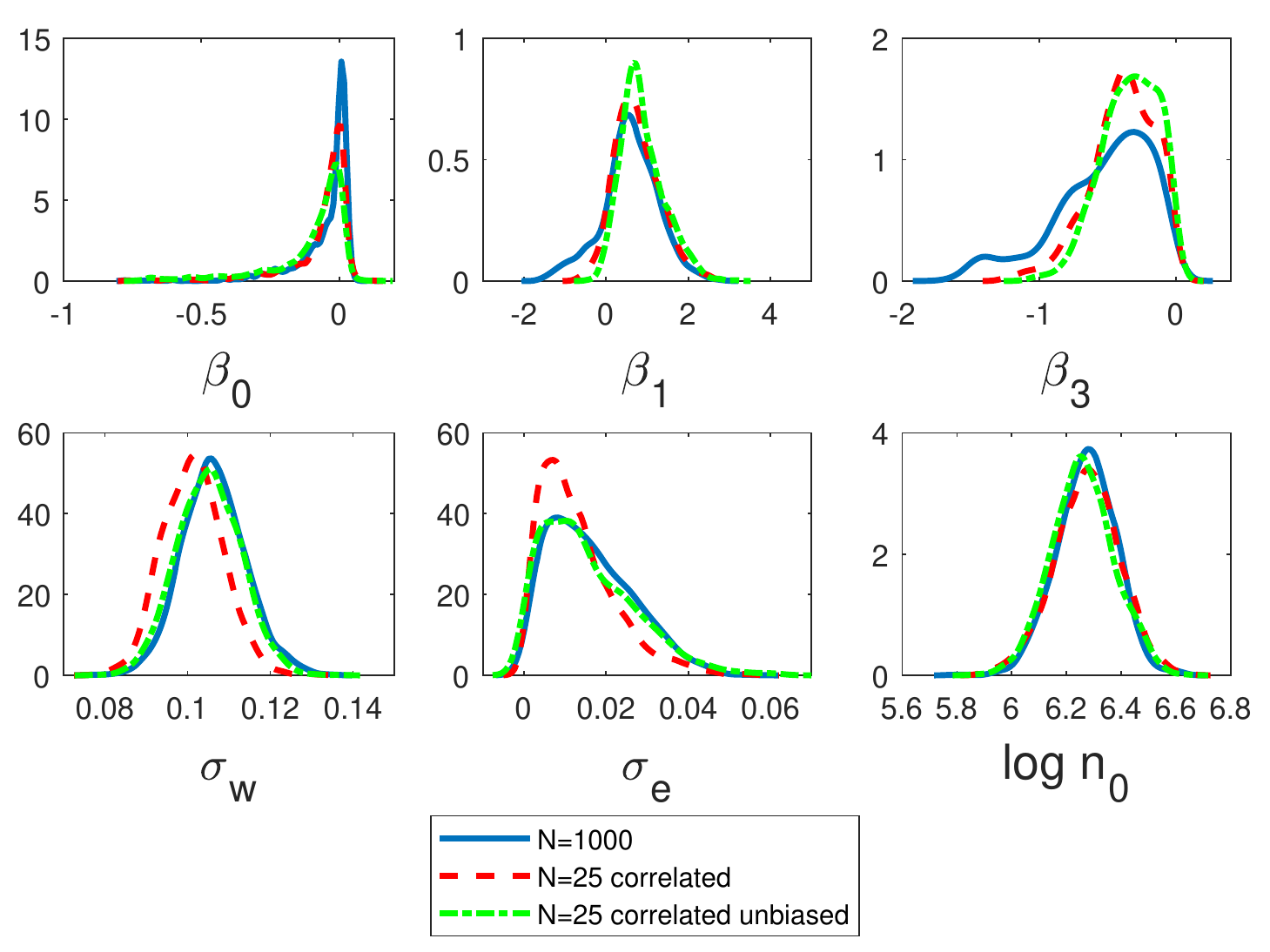}
	\end{center}
	\caption{Approximate univariate posteriors for the Ricker model using eMCMC with $N=1000$ (blue solid), correlated eMCMC with $N=25$ (red dash) and correlated ueMCMC with $N=25$ (green dash-dot).}
	\label{fig:ecol_posteriors_thetalog_correlated}
\end{figure}

\begin{figure}[tbhp]
	\begin{center}
		\includegraphics[width=0.7\textwidth]{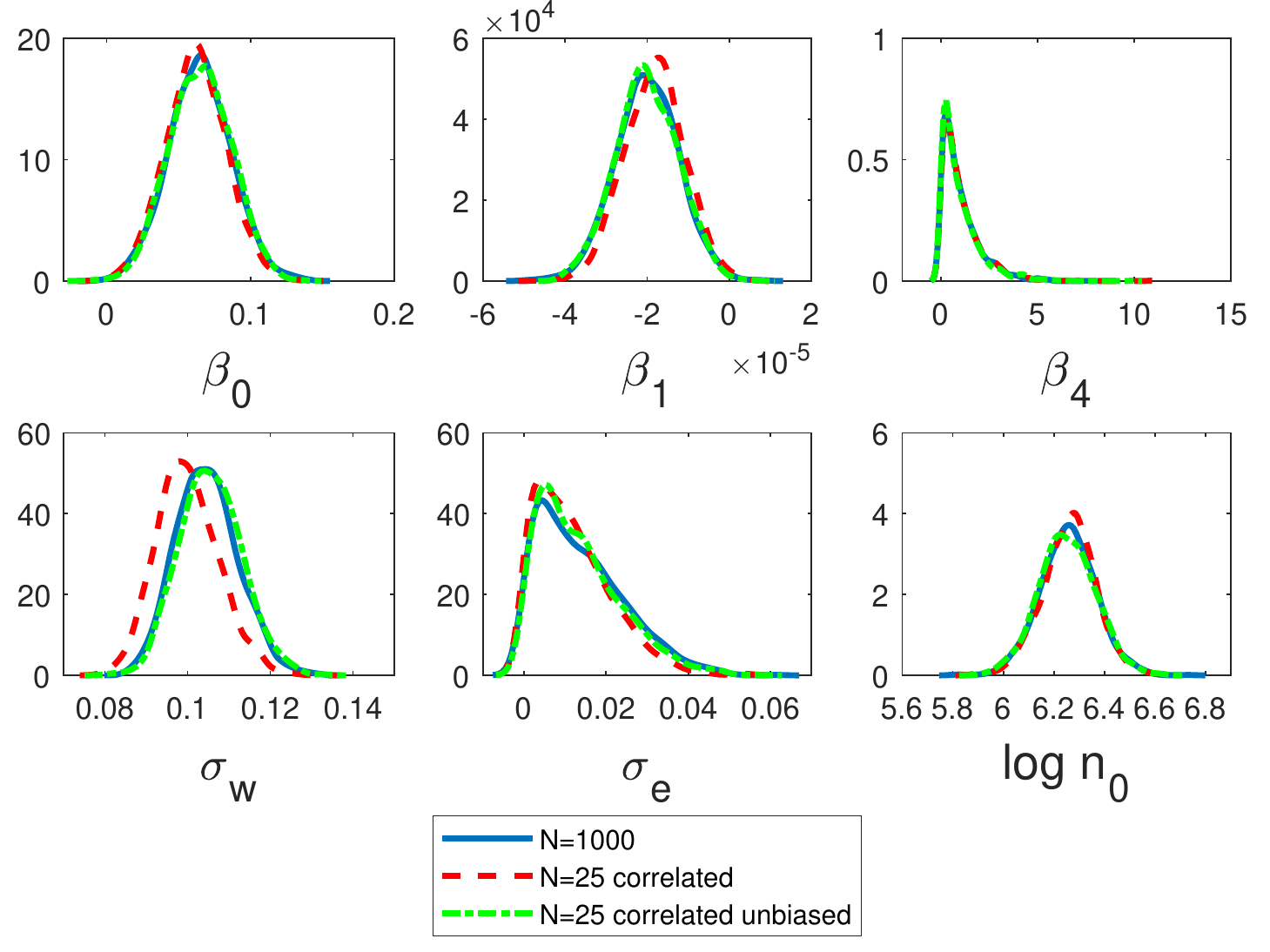}
	\end{center}
	\caption{Approximate univariate posteriors for the Ricker model using eMCMC with $N=1000$ (blue solid), correlated eMCMC with $N=25$ (red dash) and correlated ueMCMC with $N=25$ (green dash-dot).}
	\label{fig:ecol_posteriors_matelimited_correlated}
\end{figure}

\begin{figure}[tbhp]
	\begin{center}
		\includegraphics[width=0.7\textwidth]{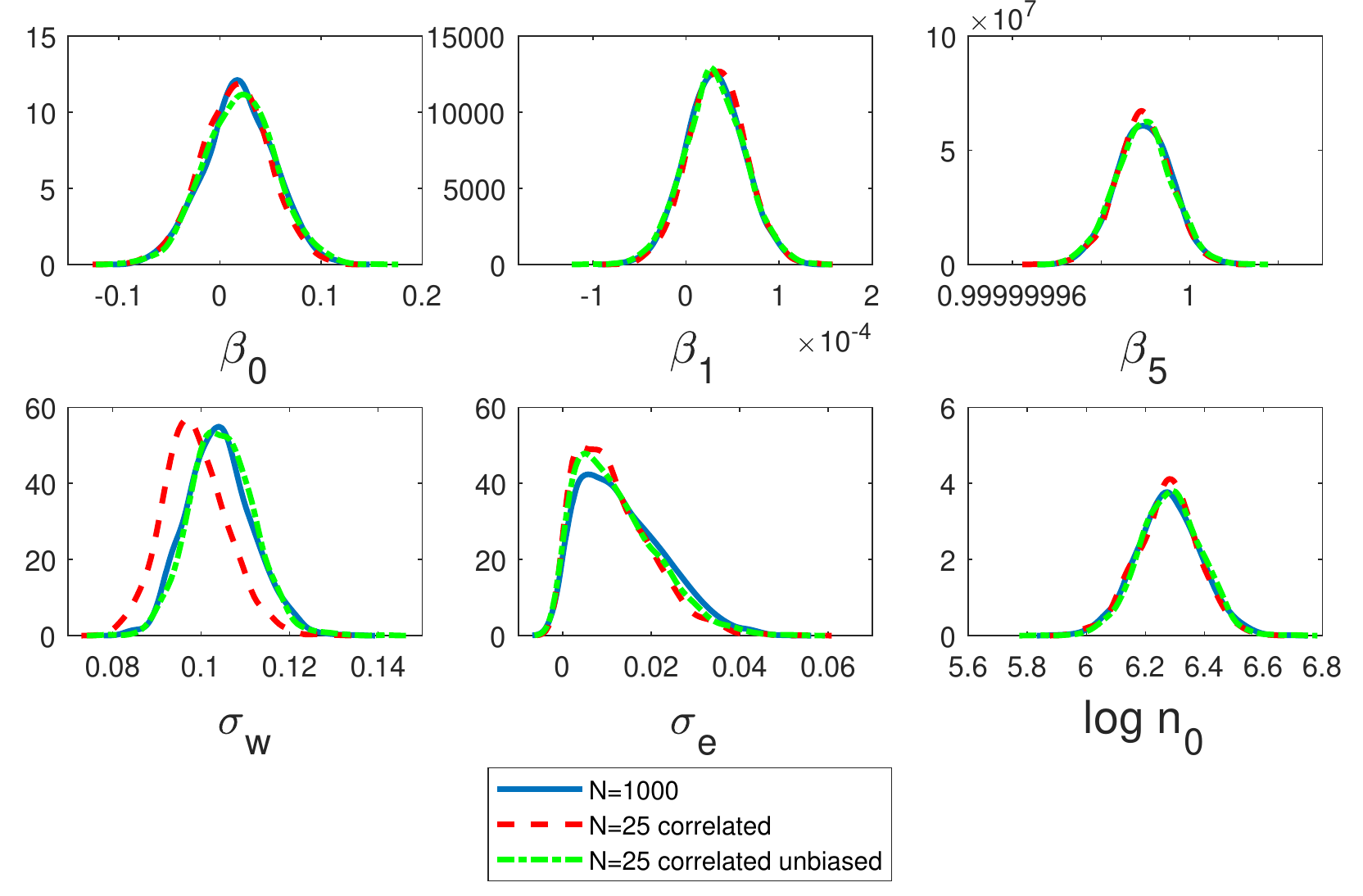}
	\end{center}
	\caption{Approximate univariate posteriors for the Ricker model using eMCMC with $N=1000$ (blue solid), correlated eMCMC with $N=25$ (red dash) and correlated ueMCMC with $N=25$ (green dash-dot).}
	\label{fig:ecol_posteriors_flexricker_correlated}
\end{figure}

\clearpage

\section*{Appendix B - RQMC results for the population ecology example} \label{app:rqmc_results}

\begin{figure}[tbhp]
	\begin{center}
		\includegraphics[width=0.7\textwidth]{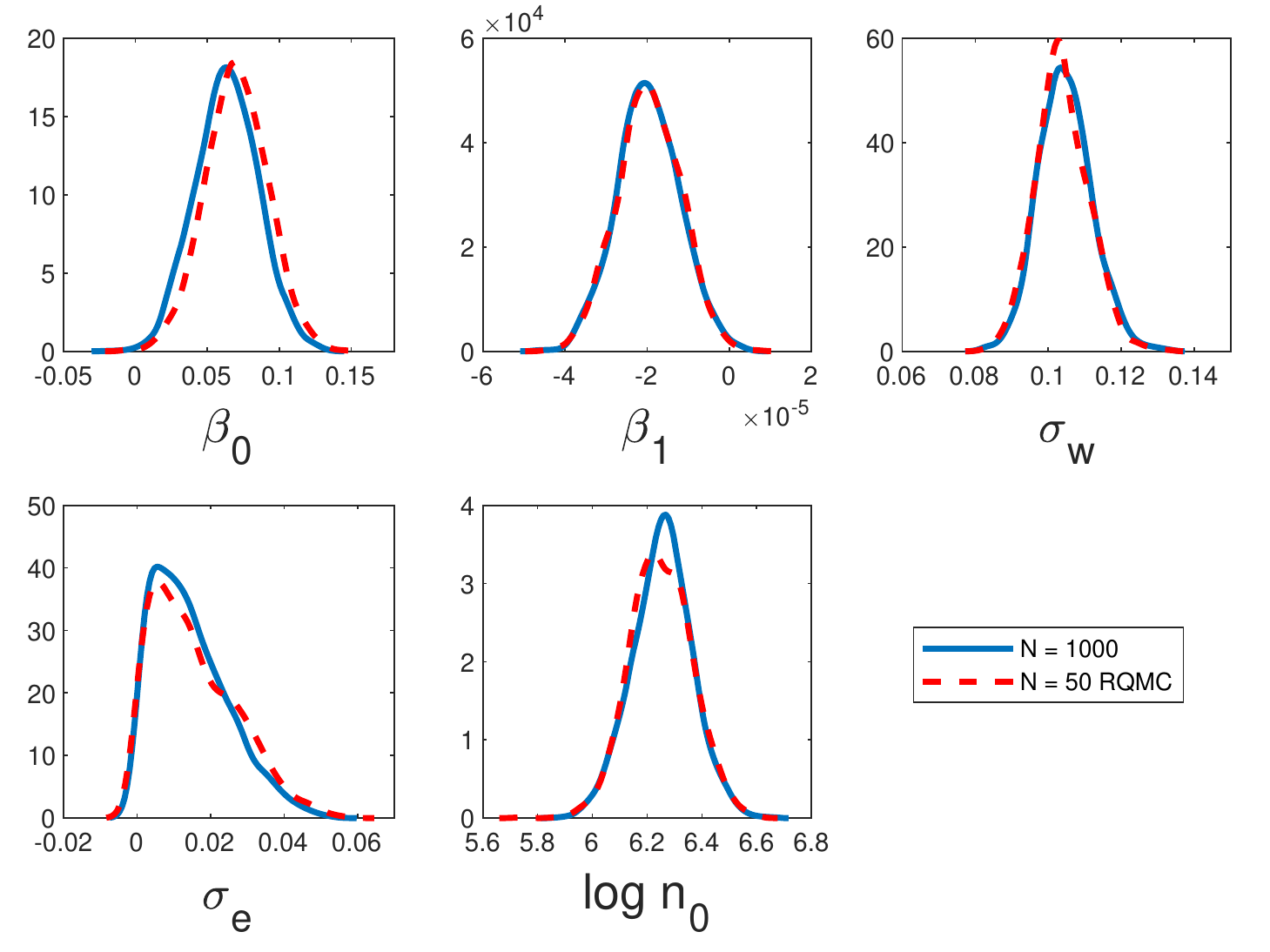}
	\end{center}
	\caption{Approximate univariate posteriors for the Ricker model using eMCMC with $N=1000$ (blue solid) and rqmc eMCMC with $N=50$ (red dash).}
	\label{fig:ecol_posteriors_ricker_rqmc}
\end{figure}

\begin{figure}[tbhp]
	\begin{center}
		\includegraphics[width=0.7\textwidth]{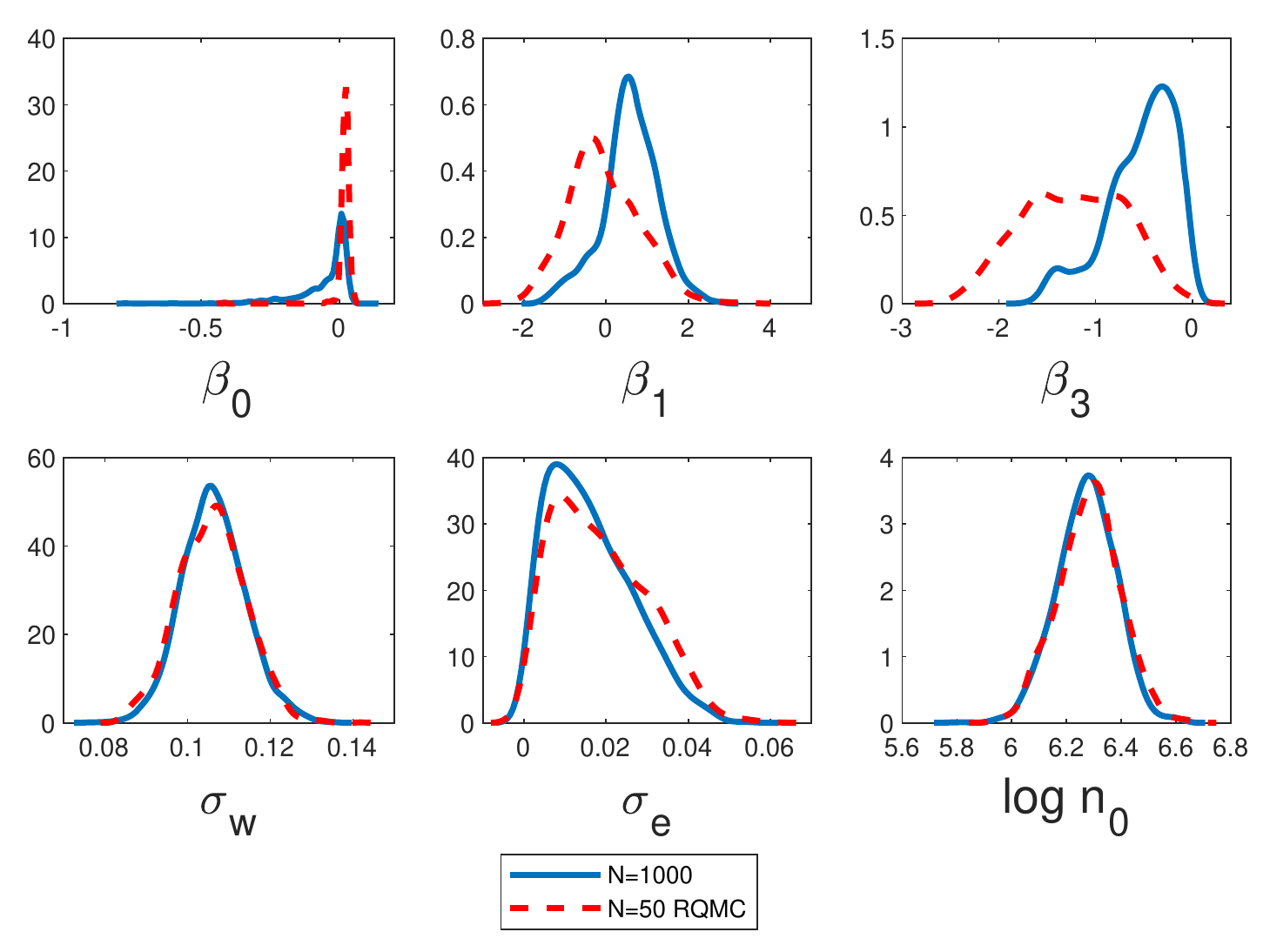}
	\end{center}
	\caption{Approximate univariate posteriors for the theta-logistic model using eMCMC with $N=1000$ (blue solid) and rqmc eMCMC with $N=50$ (red dash).}
	\label{fig:ecol_posteriors_thetalog_rqmc}
\end{figure}

\begin{figure}[tbhp]
	\begin{center}
		\includegraphics[width=0.7\textwidth]{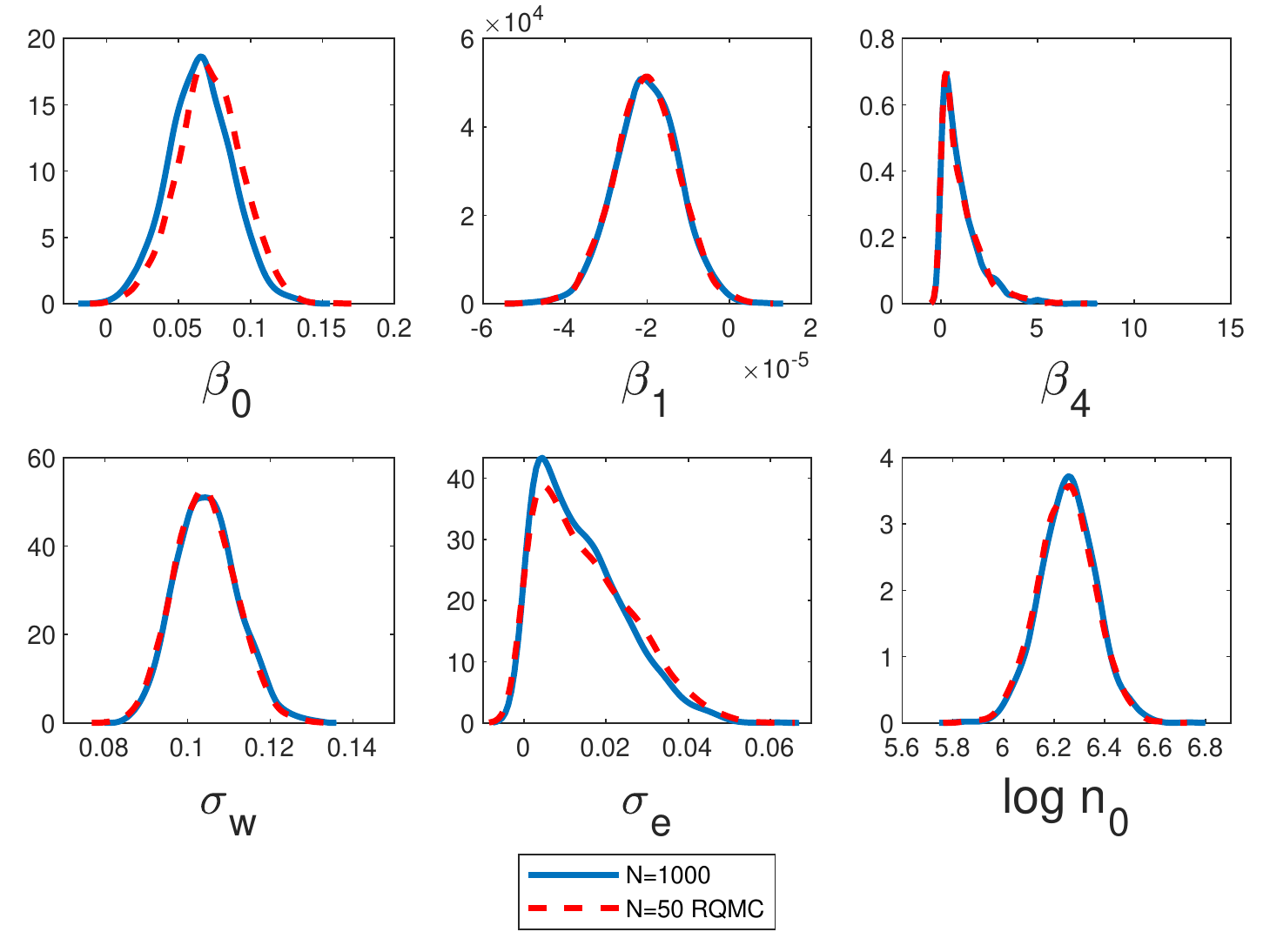}
	\end{center}
	\caption{Approximate univariate posteriors for the mate-limited model using eMCMC with $N=1000$ (blue solid) and rqmc eMCMC with $N=50$ (red dash).}
	\label{fig:ecol_posteriors_matelimited_rqmc}
\end{figure}

\begin{figure}[tbhp]
	\begin{center}
		\includegraphics[width=0.7\textwidth]{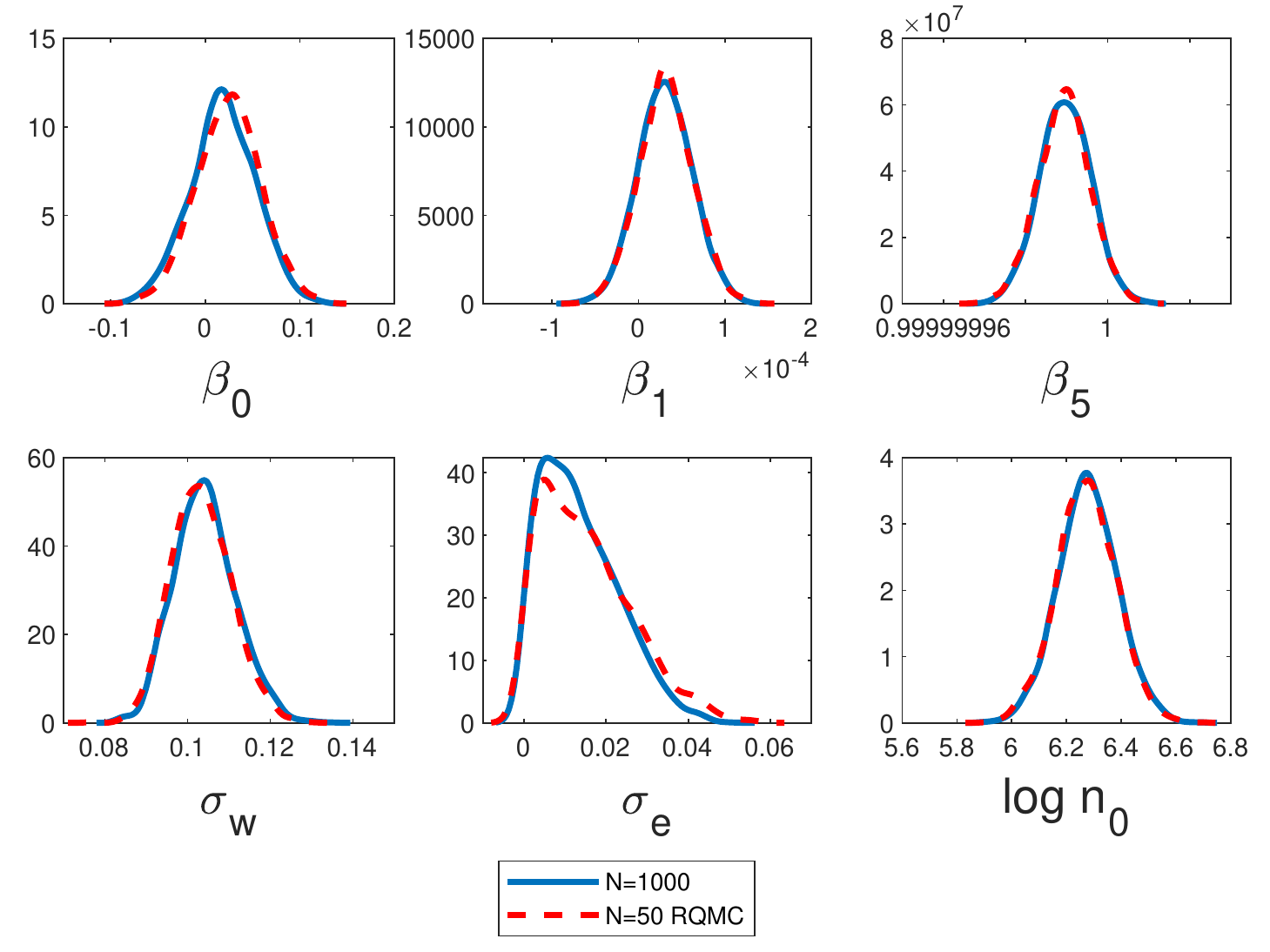}
	\end{center}
	\caption{Approximate univariate posteriors for the theta-logistic model using eMCMC with $N=1000$ (blue solid) and rqmc eMCMC with $N=50$ (red dash).}
	\label{fig:ecol_posteriors_flexricker_rqmc}
\end{figure}

\end{document}